\documentclass[10pt,journal,compsoc]{IEEEtran}
\usepackage{graphicx}
\usepackage{amsmath,amsthm}
\usepackage{amsmath,bm}
\usepackage{amsfonts}
\usepackage{cases}
\usepackage{booktabs}
\usepackage{subfigure}

\ifCLASSOPTIONcompsoc
  \usepackage[nocompress]{cite}
\else
  \usepackage{cite}
\fi
\ifCLASSINFOpdf
\else
\fi
\begin{document}
%
\title{Computation Offloading in Heterogeneous Mobile Edge Computing with Energy Harvesting}
%
%
%
%

\author{Tian~Zhang,
        and~ Wei~Chen,~\IEEEmembership{Senior~Member,~IEEE}
\IEEEcompsocitemizethanks{\IEEEcompsocthanksitem T. Zhang is with the School of Information Engineering, Shandong Management University, Jinan 250357, China.\protect\\
E-mail: tianzhang.ee@gmail.com.
\IEEEcompsocthanksitem W. Chen is with the Department of Electronic Engineering, Tsinghua University, Beijing 100084, China.\protect\\
E-mail: wchen@tsinghua.edu.cn.
}
}

\IEEEtitleabstractindextext{%
\begin{abstract}
Energy harvesting aided mobile edge computing (MEC) has gained much attention for its widespread application in the computation-intensive, latency-sensitive and energy-hungry scenario. Computation offloading, which leverages powerful MEC servers (MEC-ss) to augment the computing capability of less powerful mobile devices (MDs), is intrinsically a distributed computing over heterogeneous MEC networks. In this paper, computation offloading from multi-MD to multi-MEC-s in heterogeneous MEC systems with energy harvesting is investigated from a game theoretic perspective. The objective is to minimize the average response time of an MD that consists of communication time, waiting time and processing time. $M/G/1$ queueing models are established for MDs and MEC-ss. The interference among MDs, the randomness in computation task generation, harvested energy arrival, wireless channel state, queueing at the MEC-s, and the power budget constraint of an MD are taken into consideration. A noncooperative computation offloading game is formulated. The action is a vector that denotes the amount of computation tasks offloaded to all MEC-ss (the element value can be zero) and local process. We give the definition and existence analysis of the Nash equilibrium (NE). Furthermore, we reconstruct the optimization problem of an MD. A 2-step decomposition is presented and performed. Thereby, we arrive at a one-dimensional search problem and a greatly shrunken sub-problem. The sub-problem is nonconvex, but its Karush-Kuhn-Tucker (KKT) conditions have finite solutions. We can obtain the optimal solution of the sub-problem by seeking the finite solutions. Thereafter, a distributive NE-orienting iterated best-response algorithm is designed. Simulations are carried out to illustrate the convergence performance and parameter effect of the proposed algorithm.
\end{abstract}

\begin{IEEEkeywords}
Mobile edge computing, computation offloading, energy harvesting, game theory, queueing theory, Karush-Kuhn-Tucker condition.
\end{IEEEkeywords}}

\maketitle

\IEEEdisplaynontitleabstractindextext

%
\IEEEpeerreviewmaketitle

\IEEEraisesectionheading{\section{Introduction}}

%
%
%
%
\IEEEPARstart{M}{obile} edge computing (MEC) that provides cloud services within the vicinity of mobile device (MD) via the radio access network can effectively enhance the computing power of MD and reduce latency\cite{N. Abbas Y. Zhang A. Taherkordi and T. Skeie201802:IEEETTJ}-\cite{T Zhang 201802 IEEEAccess}. Computation offloading, which leverages powerful MEC servers to augment the computing capability of less powerful MDs, is a key computing paradigm used in MEC. In \cite{Keqin Li:IEEETSUSC}, computation offloading strategy optimization with
multiple heterogeneous servers in MEC is investigated. Three multi-variable optimization problems are studied, and corresponding efficient numerical algorithms are proposed. In \cite{L. Dong M. N. Satpute J. Shan B. Liu Y. Yu T. Yan201907:IEEEICDCS}, the computation offloading problem is formulated as graph cut problem, and
 a solution based on spectral clustering computation is developed.
In \cite{J. L. D. Neto201811:IEEETMC}, a lightweight and efficient user level online offloading framework for MEC, is presented. Real experiments with Android systems and simulations using large-scale data from a cellular network provider are preformed to prove the efficiency of the proposed framework. In \cite{X. Chen L. Jiao W. Li X. Fu201610:IEEEToN}, the multi-user computation offloading problem for mobile-edge cloud computing is studied under a multi-channel wireless interference environment. A multi-user computation offloading game is formulated, and  a distributed Nash equilibrium (NE) achieving computation offloading algorithm is designed.

To prolong the lifetime of the battery, energy harvesting technology being capable of converting the energy from the environment (e.g., solar, ambient radio-frequency (RF) signals) into electrical energy has been widely utilized in wireless communications\cite{S. Ulukus A. Yener E. Erkip O. Simeone M. Zorzi P. Grover and K. Huang201501:IEEE JSAC}\cite{T. Zhang201504:IEEE Trans. Veh. Technol.}. The energy management of the stored harvested energy at the battery plays a centric rule in energy harvested aided wireless networks. In \cite{H. Azarhava J. M. Niya:IEEE WCL}, energy efficient resource allocation in a  time division multiple access (TDMA) based wireless energy harvesting sensor network is studied. The closed form expression for the optimization problem is obtained, and thereafter solved by utilizing Karush-Kuhn-Tucker (KKT) conditions. In \cite{D. Altinel and G. K. Kurt201906:IEEE Transactions on Green Communications and Networking}, hybrid energy harvesting communication systems are modeled based on their probabilistic natures. An integrated Markov energy model is designed. In \cite{Y. Alsaba2018:IEEE Communications Surveys Tutorials}, beamforming implementation in energy harvesting wireless networks is surveyed. Different beamforming approaches in each network topology are classified according to its design objective.

Over the past years, massive multi-player online game, virtual reality (VR), and artificial intelligence (AI)-based applications (e.g., augmented reality and face detection) have emerged on smart MDs. High computation capacity requirement, low latency, heavy energy consumption, and the stringent equipment-size constraint have brought great challenges on smart MDs, especially for the CPU and battery. Combing the virtues of the two aforementioned technologies, energy harvesting MEC could augment the MD to handle the computation-intensive, latency-sensitive and energy-hungry applications.
In \cite{Y. Mao J. Zhang and K. B. Letaief201612:IEEE JSAC}, a green MEC system with energy harvesting devices has been considered, and an effective Lyapunov optimization-based dynamic computation offloading algorithm is developed. The proposed algorithm jointly decides the offloading decision, the CPU-cycle frequencies for mobile execution, and the transmit power for computation offloading. In \cite{J. Xu L. Chen and S. Ren201707:IEEE TCCN}, an efficient
reinforcement learning-based resource management algorithm is designed for energy harvesting MEC. The proposed algorithm learns on-the-fly the optimal policy of dynamic workload
offloading to the centralized cloud and edge server provisioning to minimize the long-term system cost (including both service delay and operational cost). In \cite{Z. Wei201906:IEEE itj}, the computation offloading process in energy harvesting MEC is modeled as a Markov decision process (MDP) with no prior statistic information. Thereafter, reinforcement learning algorithms are utilized to derive the optimal offloading policy.
In \cite{W. Chen D. Wang and K. Li201909:IEEE TSC}, the multi-user multi-task computation offloading problem for green MEC is investigated. Lyaponuv optimization approach is utilized to determine the energy harvesting policy and the task offloading schedule. Centralized and distributed greedy maximal scheduling algorithms are proposed.
In \cite{D. Zeng201911:IEEE netw}, the energy harvesting edge computing is investigated, and an effective and efficient MDP-based energy harvesting and data transmission scheduling strategy is designed.
In \cite{D. Zhang202004:IEEE TMC}, computation offloading is discussed in an edge-computing system consisting of energy harvesting MDs and a dispatcher. Online rewards-optimal auction is developed to maximize the long-term sum-of-rewards for processing offloaded tasks.
In \cite{F. Zhou R. Q.Hu:IEEE TWC}, computation efficiency maximization problems are investigated in wireless-powered MEC networks. Partial and binary computation offloading modes for TDMA and non-orthogonal multiple access (NOMA) are respectively considered. In \cite{L. Huang S. Bi and Y.-J. A. Zhang:IEEE TMC}, deep reinforcement learning-based online offloading framework to maximize
the weighted sum computation rate in wireless powered
MEC networks with binary computation offloading is proposed. The proposed algorithm optimally adapts wireless resource allocations and task offloading decisions to the time-varying wireless channel conditions.

In this paper, we investigate the computation offloading in multiple heterogeneous MEC networks with energy harvesting. There are multiple MDs and multiple MEC serves (MEC-ss) from different MEC networks. Besides tackling locally, each energy harvesting aided MD could partially or fully offload its computation tasks to MEC-ss via wireless channels.
There are interferences when MDs offload tasks to MEC-ss simultaneously over the shared wireless channel. Thereby the data transmission time is influenced for each MD.
We settle $M/G/1$ queueing models for MDs and MEC-ss. Tasks from different MDs wait in a queue at each MEC-s. Therefore the waiting time is affected for each other. That is to say, MDs' offloading strategies have impact on response time of each other.
For each MD, the offloading should consider physical conditions: The computation task generation, the harvested energy arrival, the wireless channel state, the queueing time, and the power budget constraint. \emph{How to choose MEC-ss and how many should be offloaded} becomes a challenging problem for each MD. First, a noncooperative game framework is established to describe the MDs' offloading. In the game, MDs are players and the action is a vector that corresponds to the offloading amount of tasks for each MEC-s (the value is 0 for these MEC-ss that are not selected for offloading) and for local process. There are constraints on actions to comply with physical conditions. The payoff is the average response time of computation offloading for each MD that includes the communication time, the computing time, and the waiting time. Next, the best-response iterated algorithm is designed for the proposed game. Although the optimization problem of each MD is intractable, it is decomposed into a 2-step structure: one dimensional search and a sub-problem. The sub-problem is NOT convex, but it can be solved through searching over the finite solutions of KKT conditions.

In conclusion, the contributions of the paper are in three-fold.
\begin{itemize}
\item We consider multiple heterogeneous energy harvesting MEC systems with multiple MDs and multiple MEC-ss. The randomness of computation task generation, harvested energy arrival, wireless channel state, queueing at the MEC-s, the power budget constraint, and the interference among MDs when offloading are taken into account.
\item A game theoretic framework is developed. We formulated a noncooperative computation offloading game for MDs, an action vector is designed to denote the offloading amount for these chosen MEC-ss. In addition, the NE is defined and its existence is analyzed.
\item The one-MD optimization problem is solved by a 2-step decomposition and KKT conditions. Thereafter, a NE-orienting distributive iterated best-response algorithm is derived of the game. Numerical results demonstrate the convergence and parameter effect of the proposed algorithm.
\end{itemize}

The rest of the paper is structured as follows. The system model is presented in Section \ref{Sec:System Model and Problem Formulation}, and we formulate the noncooperative game. In Section \ref{Sec:Problem analysis}, we carry out analysis on the optimization problem of one MD and the game. Next, the NE-orienting iterated algorithm is proposed in Section \ref{Sec:Algorithm design}. We show simulations and numerical results in Section \ref{Sec:Numerical results}. Finally, Section \ref{Sec:conclusion} concludes the whole paper.

\section{System Model and Problem Formulation}\label{Sec:System Model and Problem Formulation}
\begin{figure}[]
\centering
\includegraphics[width=3.5in]{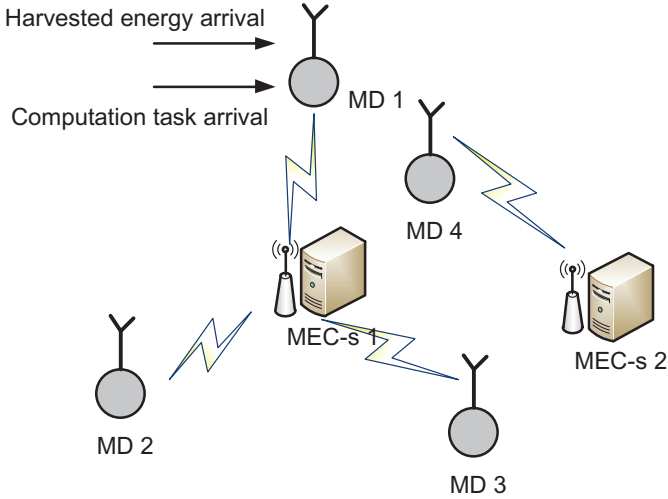}
\caption{Heterogeneous MEC systems with multi-MD and multi-MEC-s.}
\label{fig_system model}
\end{figure}
As shown in Fig. \ref{fig_system model}, consider multiple heterogeneous MEC networks consisting of $M$ MDs and $N$ MEC-ss. The MD is denoted as MD $1$, $\cdots$, MD $M$. The MEC-s is denoted as MEC-s $1$, $\cdots$, MEC-s $N$. Formally, $\mathbb{M}=\{0,\cdots,M\}$ is the MD set, and $\mathbb{N}=\{1,\cdots,N\}$ represents the MEC-s set.
Each MD is equipped with energy harvesting component that can extract energy from the environment (e.g., light, thermal, kinetic, magnetic field sources). Generally, the computation capability of the MEC-s surpasses that of the MD, thereby the MD prefers to offloading computation tasks to the MEC-s. An MD chooses an MEC-s for computation offloading in the light of some merits, e.g., the real-time vicinity, the wireless channel state, etc. Denote that MD $i$ offloads computation tasks to MEC-s $j$ with probability $p_{i,j}$. Apparently, $\sum\limits_{j\in \mathbb{N}}{p_{i,j}}=1$.
We suppose that the MDs share the same wireless frequency bandwith for computation offloading. Consider $M/G/1$  model for an MD, and assume that the computation tasks generate at MD $i$ according to a Poisson process with arrival rate $\lambda_i$. Furthermore, the stream of generated computation tasks is divided into the offloaded computation task sub-stream with rate $\tilde{\lambda}_i$ and local processing sub-stream with rate $\hat{\lambda}_i$, where $\lambda_i=\tilde{\lambda}_i+\hat{\lambda}_i$. The computation tasks offloaded from MD $i$ to MEC-s $j$ constitute a Poisson stream with arrival rate $\tilde{\lambda}_{i,j}=p_{i,j}\tilde{\lambda}_i$. For MEC-s $j$, the received computation tasks from MDs compose a Poisson stream with arrival rate $\pi_j=\sum\limits_{i\in \mathbb{M}}\tilde{\lambda}_{i,j}$. $M/G/1$ queueing model is erected for each MEC-s.
Denote the execution requirements (e.g., number of processor cycles to be executed) of computation tasks generated at MD $i$ as independent and identically distributed (i.i.d.) random variable $r_i$. Let $f_i$ and $F_j$ be the processor's computing speed (cycles per second, i.e., HZ) of MD $i$ and MEC-s $j$, respectively.
The amount of data to be communicated between MD $i$ and MEC-s $j$ is depicted as i.i.d. random variables $A_{i,j}$. The data transmission rate between MD $i$ and MEC-s $j$ is $R_{i,j}$.
The time of locally processing at MD $i$ is $\frac{r_i}{f_i}$. The processing time of offloaded computation tasks from MD $i$ to MEC-s $j$ is $\frac{r_i}{F_j}+\frac{A_{i,j}}{R_{i,j}}$, where $\frac{r_i}{F_j}$ is the computing time and $\frac{A_{i,j}}{R_{i,j}}$ is the data transmission time. The processing time of computation tasks at MEC-s $j$ can be characterized by random variable $\tau_j$ with mean
\begin{eqnarray}\label{mean of Processing time at MEC-s j}
\bar{\tau_j}=\sum_{i\in\mathbb{M}}p_{i,j}\left(\frac{\bar{r_i}}{F_j}+\frac{\bar{A_{i,j}}}{R_{i,j}}\right)
\end{eqnarray}
and second moment
\begin{eqnarray}\label{Second moment of Processing time at MEC-s j}
\bar{\tau_j^2}=\sum_{i\in\mathbb{M}}p_{i,j}\left(\frac{\bar{r_i^2}}{F_j}+\frac{2\bar{r_i}\bar{A_{i,j}}}{F_j R_{i,j}}+\frac{\bar{A_{i,j}^2}}{R_{i,j}}\right).
\end{eqnarray}
The average waiting time of computation tasks at MEC-s $j$ can be given by \cite{Kleinrock1975:Queueing Systems Volume 1}
\begin{eqnarray}\label{Waiting time at MEC-s j}
\omega_j=\frac{\pi_j\bar{\tau_j^2}}{2(1-\zeta_j)},
\end{eqnarray}
where $\zeta_j=\pi_j\bar{\tau_j}$ is the utilization of MEC-s $j$.
The average response time of offloaded computation tasks from MD $i$ to MEC-s $j$ can be expressed as
\begin{eqnarray}\label{responsing time from MD i to MEC-s j}
T_{i,j}=\frac{\bar{r_i}}{F_j}+\frac{\bar{A_{i,j}}}{R_{i,j}}+\omega_j
\end{eqnarray}
The average response time of generated computation tasks on MD $i$ is given by
\begin{eqnarray}\label{Average responsing time at MD i}
T_{i}=\frac{\hat{\lambda}_i}{\lambda_i}\frac{\bar{r_i}}{f_i}+\frac{\tilde{\lambda}_i}{\lambda_i}\sum_{j\in\mathbb{N}}p_{i,j}T_{i,j}.
\end{eqnarray}
The channel coefficient between MD $i$ and MEC-s $j$ is $h_{i,j}$, the channel bandwidth $B$, the transmission power from MD $i$ to MEC-s $j$ is $P_{i,j}$.
The data transmission rate is given by
$R_{i,j}=B \log_2(1+\frac{P_{i,j}|h_{i,j}|^2}{N_0+I_j})$, where $N_0$ is the noise power and $I_j=\sum_{\substack{l\ne i, l\in \mathbb{M}\\k\in\mathbb{N}}}P_{l,k}|h_{l,j}|^2$ is the received interference.
The power consumption can be expressed by
\begin{eqnarray}\label{Rate power relation}
P_{i,j}=\xi_{i,j}\left(2^{\frac{R_{i,j}}{B}}-1\right),
\end{eqnarray}
where $\xi_{i,j}=\frac{N_0+I_j}{|h_{i,j}|^2}$.
The average energy consumption of data transmission from MD $i$ to MEC-s $j$ for a computation task offloading is $P_{i,j}\frac{\bar{A_{i,j}}}{R_{i,j}}$.
The average data transmission energy consumption for a computation task offloading at MD $i$ can be written as
\begin{eqnarray}
E_{t,i}=\sum_{j\in\mathbb{N}}p_{i,j}P_{i,j}\frac{\bar{A_{i,j}}}{R_{i,j}}.
\end{eqnarray}
Power consumption due to computation at MD $i$ can be given by
$\eta_i f_i^3$, where $\eta_i$ is the computation energy efficiency\cite{S. Guo201604:Proc. IEEE INFOCOM}. The average computation power consumption at MD $i$ can be described as
\begin{eqnarray}
P_{c,i}=\sigma_i\eta_i f_i^3+P_s,
\end{eqnarray}
where $\sigma_i=\hat{\lambda_i}\frac{\bar{r_i}}{f_i}$ is the processor utilization of MD $i$, $P_s$ is the power consumption when no computation is executed.
The average power consumption of communication and computation at MD $i$ can be expressed as
\begin{eqnarray}
P_i=\tilde{\lambda}_iE_{t,i}+P_{c,i}
\end{eqnarray}
Assume that the harvested energy arrival of MD $i$ is i.i.d with rate $\mathcal{E}_i$. The greedy strategy of harvested energy usage, i.e., utilizing the available harvested energy preferentially, is optimal with high probability\cite{T. Zhang201504:IEEE Trans. Veh. Technol.}. Therefore, the average extra power that is constrained can be written as $P_i-\mathcal{E}_i$.

MD $i$ aims to minimize the average response time under the power constraint. Mathematically,
\begin{eqnarray} \label{minimization problem of MD i}
\mathrm(P_1)~\min_{\left\{\tilde{\lambda}_{i,j}\right\}_{j\in \mathbb{N}}} T_i
\end{eqnarray}
\begin{subequations}
\begin{numcases}{ \mbox{s.t.}}
\tilde{\lambda}_{i,j}\ge0, \forall j\in\mathbb{N},\\
\tilde{\lambda}_{i}\le \lambda_i,\\
\zeta_j\le 1,\forall j \in \mathbb{N},\label{Processor utilization constraint}\\
P_i-\mathcal{E}_i\le \mathfrak{C}_i.\label{power constraint}
\end{numcases}
\end{subequations}
where $\mathfrak{C}_i$ is the power constraint on MD $i$.
MDs are self-interested and compete each other to minimize its own average response time by adjusting the computation offloading strategy. Formally, the noncooperative game of MDs can be formulated as
\begin{eqnarray}
\mathcal{G}=\left\{\Omega,\left\{\mathcal{S}_i\right\}_{i\in \Omega},\left\{\mathcal{U}_i\right\}_{i\in \Omega}\right\}
\end{eqnarray}
where $\Omega=\mathbb{M}$ is the set of players,
\begin{eqnarray}
\mathcal{S}_i=\Big\{\textbf{S}_i=\left(\tilde{\lambda}_{i,1},\cdots,\tilde{\lambda}_{i,N}\right):\tilde{\lambda}_{i,j}\ge0, \forall j\in \mathbb{N}, \nonumber\\
\tilde{\lambda}_{i}\le \lambda_i, \zeta_j\le 1, \forall j\in \mathbb{N},P_i-\mathcal{E}_i\le \mathfrak{C}_i.\Big\}
\end{eqnarray}
is the offloading strategy\footnote{In the paper, we focus on the computation allocation strategy. When the data transmission rate $\{R_{i,j}\}_{j=1,\cdots,N}$ is added in the strategy, it corresponds to the joint allocations of computation and resource.} of player $i$, and $\mathcal{U}_i=T_i$ is the payoff.

\section{Problem analysis}\label{Sec:Problem analysis}
\theoremstyle{lemma} \newtheorem{lemma}{Lemma}
\theoremstyle{definition} \newtheorem{definition}{Definition}
\theoremstyle{property} \newtheorem{property}{Property}
Optimization problem of an MD is analyzed in Section \ref{Sec:Optimization problem of one player}.
To analyze and handle the formulated problem conveniently, we first reconstruct the objective function and the constraints. The corresponding KKT conditions are derived. However, it is very difficult to solve. Next, a 2-step decomposition strategy is built, and we further reexpress the optimization problem accordingly. The optimization can be decomposed into a one-dimension search together with a reduced subproblem. In Section \ref{Sec:Game analysis}, game analysis is performed. The definition and existence of mixed strategy NE are investigated.

\subsection{Optimization problem of MD $i$}\label{Sec:Optimization problem of one player}
Combing (\ref{mean of Processing time at MEC-s j}, (\ref{Second moment of Processing time at MEC-s j}), (\ref{Waiting time at MEC-s j}), (\ref{responsing time from MD i to MEC-s j}), and (\ref{Average responsing time at MD i}), we have
\begin{eqnarray}
\lefteqn{T_i=\frac{\lambda_i-\sum\limits_{j=1}^{N}\tilde{\lambda_{i,j}}}{\lambda_i}\frac{\bar{r_i}}{f_i}+\frac{1}{\lambda_i}\sum\limits_{j=1}^{N}\tilde{\lambda_{i,j}}\times
}\nonumber\\
&&\left\{
\phi_{i,j}
+\frac{\left(\sum\limits_{l=1}^{M}\tilde{\lambda_{l,j}}\right)\left(\sum\limits_{k=1}^{M}\frac{\tilde{\lambda_{k,j}}\theta_{k,j}}{\sum\limits_{j=1}^{N}\tilde{\lambda_{k,j}}}\right)}{2\left[1-
\left(\sum\limits_{l=1}^{M}\tilde{\lambda_{l,j}}\right)\left(\sum\limits_{k=1}^{M}\frac{\tilde{\lambda_{k,j}}\phi_{k,j}}{\sum\limits_{j=1}^{N}\tilde{\lambda_{k,j}}}\right)\right]}\right\},\nonumber\\
\end{eqnarray}
where $\theta_{k,j}=\frac{\bar{r_k^2}}{F_j}+\frac{2\bar{r_k}\bar{A_{k,j}}}{F_j R_{k,j}}+\frac{\bar{A_{k,j}^2}}{R_{k,j}}$,
$\phi_{k,j}=\frac{\bar{r_k}}{F_j}+\frac{\bar{A_{k,j}}}{R_{k,j}}$. Furthermore, $T_i$ can be rewritten as (\ref{T_i rewritten}).
\begin{figure*}[!t]
\begin{eqnarray}
\lefteqn{
T_i=\frac{\lambda_i-\sum\limits_{l=1}^{N}\tilde{\lambda_{i,l}}}{\lambda_i}\frac{\bar{r_i}}{f_i}+\frac{1}{\lambda_i}\sum\limits_{j=1}^{N}\tilde{\lambda_{i,j}}
\left\{
\phi_{i,j}
+\frac{\left(\tilde{\lambda_{i,j}}+\sum\limits_{k\ne i}\tilde{\lambda_{k,j}}\right)\left(\frac{\tilde{\lambda_{i,j}}\theta_{i,j}}{\sum\limits_{l=1}^{N}\tilde{\lambda_{i,l}}}+\sum\limits_{k\ne i}\frac{\tilde{\lambda_{k,j}}\theta_{k,j}}{\sum\limits_{l=1}^{N}\tilde{\lambda_{k,l}}}\right)}{2\left[1-\left(\tilde{\lambda_{i,j}}+\sum\limits_{k\ne i}\tilde{\lambda_{k,j}}\right)\left(\frac{\tilde{\lambda_{i,j}}\phi_{i,j}}{\sum\limits_{l=1}^{N}\tilde{\lambda_{i,l}}}+\sum\limits_{k\ne i}\frac{\tilde{\lambda_{k,j}}\phi_{k,j}}{\sum\limits_{l=1}^{N}\tilde{\lambda_{k,l}}}\right)\right]}\right\}
}\nonumber\\
%
&=&\frac{\lambda_i-\sum\limits_{l=1}^{N}\tilde{\lambda_{i,l}}}{\lambda_i}\frac{\bar{r_i}}{f_i}+\frac{1}{\lambda_i}\sum\limits_{j=1}^{N}\tilde{\lambda_{i,j}}\times
\nonumber\\
&&\left\{
\phi_{i,j}
+\frac{\theta_{i,j}\frac{\tilde{\lambda_{i,j}}^2}{\sum\limits_{l=1}^{N}\tilde{\lambda_{i,l}}}+\tilde{\lambda_{i,j}}\sum\limits_{k\ne i}\frac{\tilde{\lambda_{k,j}}\theta_{k,j}}{\sum\limits_{l=1}^{N}\tilde{\lambda_{k,l}}}+\frac{\tilde{\lambda_{i,j}}}{\sum\limits_{l=1}^{N}\tilde{\lambda_{i,l}}}\theta_{i,j}\left(\sum\limits_{l\ne i}\tilde{\lambda_{l,j}}\right)+\left(\sum\limits_{l\ne i}\tilde{\lambda_{l,j}}\right)\left(\sum\limits_{k\ne i}\frac{\tilde{\lambda_{k,j}}\theta_{k,j}}{\sum\limits_{l=1}^{N}\tilde{\lambda_{k,l}}}\right)}
{2\left[1-\left(\phi_{i,j}\frac{\tilde{\lambda_{i,j}}^2}{\sum\limits_{l=1}^{N}\tilde{\lambda_{i,l}}}+\tilde{\lambda_{i,j}}\sum\limits_{k\ne i}\frac{\tilde{\lambda_{k,j}}\phi_{k,j}}{\sum\limits_{l=1}^{N}\tilde{\lambda_{k,l}}}+\frac{\tilde{\lambda_{i,j}}}{\sum\limits_{l=1}^{N}\tilde{\lambda_{i,l}}}\phi_{i,j}\left(\sum\limits_{l\ne i}\tilde{\lambda_{l,j}}\right)+\left(\sum\limits_{l\ne i}\tilde{\lambda_{l,j}}\right)\left(\sum\limits_{k\ne i}\frac{\tilde{\lambda_{k,j}}\phi_{k,j}}{\sum\limits_{l=1}^{N}\tilde{\lambda_{k,l}}}\right)\right)\right]}\right\}
\label{T_i rewritten}
\end{eqnarray}
\end{figure*}
 Let $\alpha_{i,j}=\sum\limits_{k\ne i}\frac{\tilde{\lambda_{k,j}}\theta_{k,j}}{\sum\limits_{l=1}^{N}\tilde{\lambda_{k,l}}}$, $\beta_{i,j}=\theta_{i,j}\left(\sum\limits_{l\ne i}\tilde{\lambda_{l,j}}\right)$, $\gamma_{i,j}=\left(\sum\limits_{l\ne i}\tilde{\lambda_{l,j}}\right)\left(\sum\limits_{k\ne i}\frac{\tilde{\lambda_{k,j}}\theta_{k,j}}{\sum\limits_{l=1}^{N}\tilde{\lambda_{k,l}}}\right)$, $\delta_{i,j}=\sum\limits_{k\ne i}\frac{\tilde{\lambda_{k,j}}\phi_{k,j}}{\sum\limits_{l=1}^{N}\tilde{\lambda_{k,l}}}$, $\mu_{i,j}=\phi_{i,j}\left(\sum\limits_{l\ne i}\tilde{\lambda_{l,j}}\right)$, and $\nu_{i,j}=\left(\sum\limits_{l\ne i}\tilde{\lambda_{l,j}}\right)\left(\sum\limits_{k\ne i}\frac{\tilde{\lambda_{k,j}}\phi_{k,j}}{\sum\limits_{l=1}^{N}\tilde{\lambda_{k,l}}}\right)$. $T_i$ can be reexpressed as
\begin{eqnarray}\label{Objective function Reconstruction}
T_i&=&\frac{\lambda_i-\sum\limits_{l=1}^{N}\tilde{\lambda_{i,l}}}{\lambda_i}\frac{\bar{r_i}}{f_i}+\frac{1}{\lambda_i}\sum\limits_{j=1}^{N}\tilde{\lambda_{i,j}}\Bigg(
\phi_{i,j}
+\frac{1}{2}\times\nonumber\\
&&\frac{\theta_{i,j}\frac{\tilde{\lambda_{i,j}}^2}{\sum\limits_{l=1}^{N}\tilde{\lambda_{i,l}}}+\tilde{\lambda_{i,j}}\alpha_{i,j}+\frac{\tilde{\lambda_{i,j}}}{\sum\limits_{l=1}^{N}\tilde{\lambda_{i,l}}}
\beta_{i,j}+\gamma_{i,j}}
{1-\left(\phi_{i,j}\frac{\tilde{\lambda_{i,j}}^2}{\sum\limits_{l=1}^{N}\tilde{\lambda_{i,l}}}+\tilde{\lambda_{i,j}}\delta_{i,j}+\frac{\tilde{\lambda_{i,j}}}{\sum\limits_{l=1}^{N}\tilde{\lambda_{i,l}}} \mu_{i,j}+\nu_{i,j}\right)}
\Bigg).\nonumber\\
\end{eqnarray}

The processor utilization constraint (\ref{Processor utilization constraint}) can be expressed as
\begin{eqnarray}\label{utilization constraint Reconstruction}
\zeta_j&=&\left(\sum\limits_{l=1}^{M}\tilde{\lambda_{l,j}}\right)\left(\sum\limits_{k=1}^{M}\frac{\tilde{\lambda_{k,j}}\phi_{k,j}}{\sum\limits_{l=1}^{N}\tilde{\lambda_{k,l}}}\right)\nonumber\\
&=&\left(\tilde{\lambda_{i,j}}+\sum\limits_{l\ne i}\tilde{\lambda_{l,j}}\right)\left(\frac{\tilde{\lambda_{i,j}}\phi_{i,j}}{\sum\limits_{l=1}^{N}\tilde{\lambda_{i,l}}}+\sum\limits_{k\ne i}\frac{\tilde{\lambda_{k,j}}\phi_{k,j}}{\sum\limits_{l=1}^{N}\tilde{\lambda_{k,l}}}\right)\nonumber\\
&=&\phi_{i,j}\frac{\tilde{\lambda_{i,j}}^2}{\sum\limits_{l=1}^{N}\tilde{\lambda_{i,l}}}+\tilde{\lambda_{i,j}}\delta_{i,j}+\frac{\tilde{\lambda_{i,j}}}{\sum\limits_{l=1}^{N}\tilde{\lambda_{i,l}}}
\mu_{i,j}+\nu_{i,j}\label{Conzeta}\\
&\le&1.
\end{eqnarray}

The power constraint (\ref{power constraint}) can be written as
\begin{eqnarray}\label{power constraint Reconstruction}
P_i&=&\tilde{\lambda}_i\sum\limits_{j=1}^N\frac{\tilde{\lambda}_{i,j}}{\tilde{\lambda}_i}P_{i,j}\frac{\bar{A_{i,j}}}{R_{i,j}}+\hat{\lambda_i}\frac{\bar{r_i}}{f_i}\eta_if_{i}^3+P_s \nonumber\\
&=&\sum\limits_{j=1}^N\tilde{\lambda}_{i,j}P_{i,j}\frac{\bar{A_{i,j}}}{R_{i,j}}+\left(\sum\limits_{j=1}^N\tilde{\lambda}_{i,j}\right)\frac{\bar{r_i}}{f_i}\eta_if_{i}^3+P_s\nonumber\\
&=&\sum\limits_{j=1}^N\tilde{\lambda}_{i,j}\left(P_{i,j}\frac{\bar{A_{i,j}}}{R_{i,j}}+\frac{\bar{r_i}}{f_i}\eta_if_{i}^3\right)+P_s\label{ConP}\\
&\le& \mathcal{E}_i+ \mathfrak{C}_i.
\end{eqnarray}
$\alpha_{i,j}$,$\beta_{i,j}$, $\gamma_{i,j}$, $\delta_{i,j}$, $\phi_{i,j}$, and $\nu_{i,j}$ are constant with respect to $\{\tilde{\lambda}_{i,j}\}_{j=1,\cdots,N}$, the objective and the constraints are rewritten as the functions of $\{\tilde{\lambda}_{i,j}\}_{j=1,\cdots,N}$, explicitly and concisely.
Thereafter, the KKT conditions can be obtained in Lemma \ref{Lemma:KKT Condition}.
\begin{lemma}\label{Lemma:KKT Condition}
The KKT conditions of optimization problem $\mathrm{(P_1)}$ are
\begin{subequations}\label{KKT Condition}
\begin{numcases}{}
\frac{\partial{T_i}}{\partial{\tilde{\lambda_{i,j}}}}-\epsilon_j+\chi+\sum\limits_{j}\varpi_j\bigg[2\left(\phi_{i,j}+\delta_{i,j}\right)\tilde{\lambda_{i,j}}
\nonumber\\
+\delta_{i,j}\Big(\sum\limits_{k\ne j,k\in \mathbb{N}}\tilde{\lambda_{i,k}}\Big)+\nu_{i,j}-1+\mu_{i,j}\bigg]\nonumber\\
+ \rho\left(P_{i,j}\frac{\bar{A_{i,j}}}{R_{i,j}}+\frac{\bar{r_i}}{f_i}\eta_if_{i}^3\right)=0, \forall j\label{}\\
-\tilde{\lambda_{i,j}}\le 0,\forall j\\
\epsilon_j\ge0,\forall j\\
\epsilon_j\tilde{\lambda_{i,j}}=0,\forall j\\
  \tilde{\lambda}_{i}-\lambda_i\le0,\label{}\\
  \chi\ge 0,\\
  \chi (\tilde{\lambda}_{i}-\lambda_i)=0,\\
\zeta_j-1\le0,\label{} \forall j\\
  \varpi_j\ge0,\forall j\\
  \varpi_j\Big(\zeta_j-1\Big)=0,\forall j\\
  P_{i}-\mathcal{E}_i- \mathfrak{C}_i\le 0,\\
  \rho\ge 0, \\
  \rho(P_{i}-\mathcal{E}_i- \mathfrak{C}_i)=0 \label{}
\end{numcases}
\end{subequations}
where $\epsilon_j$, $\chi$, $\varpi_j$, $\rho$ are multipliers, $\zeta_j$ is stated by (\ref{Conzeta}), $P_{i}$ is expressed as (\ref{ConP}), and $\frac{\partial{T_i}}{\partial{\tilde{\lambda_{i,j}}}}$ is given by (\ref{Partial Ti}).
\end{lemma}
\begin{IEEEproof}
Please see Appendix \ref{Appendix:KKT proof}.
\end{IEEEproof}
The KKT conditions are necessary for the optimal solutions. Generally, (\ref{KKT Condition}) is very challenging if not intractable since the optimization variables, $\{\tilde{\lambda_{i,1}},\cdots,\tilde{\lambda_{i,N}}\}$, are coupled together.
Then we have the following lemma, i.e., Lemma \ref{Lemma:Eqivalent Optimization Problem}, to derive an equivalent optimization problem.
\begin{lemma}\label{Lemma:Eqivalent Optimization Problem}
The response time minimization problem of MD $i$, i.e., $\mathrm{(P_1)}$, can be equivalently reconstructed as
\begin{eqnarray} \label{Equivalent minimization problem of MD i}
\mathrm{(P_2)}&&\inf_{t_i\in\left[0,\lambda_i\right]}\min_{\textbf{S}_i}
\frac{\lambda_i-t_i}{\lambda_i}\frac{\bar{r_i}}{f_i}+\frac{1}{\lambda_i}\sum\limits_{j=1}^{N}\tilde{\lambda_{i,j}}\Bigg(
\phi_{i,j}
+\nonumber\\
&&\frac{1}{2}\frac{\theta_{i,j}\frac{\tilde{\lambda_{i,j}}^2}{t_i}+\tilde{\lambda_{i,j}}\alpha_{i,j}+\frac{\tilde{\lambda_{i,j}}}{t_i}
\beta_{i,j}+\gamma_{i,j}}
{1-\left(\phi_{i,j}\frac{\tilde{\lambda_{i,j}}^2}{t_i}+\tilde{\lambda_{i,j}}\delta_{i,j}+\frac{\tilde{\lambda_{i,j}}}{t_i} \mu_{i,j}+\nu_{i,j}\right)}
\Bigg)~~~~
\end{eqnarray}
\begin{subequations}\label{Equivalent constraint for minimization problem of MD i}
\begin{numcases}{\mbox{s.t.}}
\tilde{\lambda}_{i,j}\ge0, \forall j\in \mathbb{N},\\
\sum\limits_{j=1}^{N}\tilde{\lambda_{i,j}}=t_i\le \lambda_i,\\
\phi_{i,j}\frac{\tilde{\lambda_{i,j}}^2}{t_i}+\tilde{\lambda_{i,j}}\delta_{i,j}+\frac{\tilde{\lambda_{i,j}}}{t_i}
\mu_{i,j}+\nu_{i,j}-1\le0,\label{} \forall j\\
\sum\limits_{j=1}^N\tilde{\lambda}_{i,j}\left(P_{i,j}\frac{\bar{A_{i,j}}}{R_{i,j}}+\frac{\bar{r_i}}{f_i}\eta_if_{i}^3\right)+P_s-\mathcal{E}_i- \mathfrak{C}_i\le0.\label{}~~~~~~~~
\end{numcases}
\end{subequations}
\end{lemma}
\begin{IEEEproof}
Let $\sum\limits_{l=1}^{N}\tilde{\lambda_{i,l}}=t_i$. Combining (\ref{Objective function Reconstruction}), (\ref{utilization constraint Reconstruction}), (\ref{power constraint Reconstruction}) and (\ref{minimization problem of MD i}), we get (\ref{Equivalent minimization problem of MD i}) and (\ref{Equivalent constraint for minimization problem of MD i}). The proof completes.
\end{IEEEproof}
In $\mathrm{(P_2)}$, we first fix the sum of $\tilde{\lambda_{i,j}}$, i.e., $\sum\limits_{l=1}^{N}\tilde{\lambda_{i,l}}=t_i$, in the constraint. Then, we vary the sum in feasible scope. $t_i$ can be viewed as an intermediate variable. Thereby,
$\mathrm{(P_2)}$ can be tackled by a 2-step strategy. First, for a given $t_i\in \left[0,\lambda_i\right]$, solve the following shrunken problem,\footnote{The solution of ($P_3$) is investigated in Section \ref{Sec:Algorithm design}.} i.e., $\mathrm{(P_3)}$.
\begin{eqnarray} \label{minimization problem of MD i given t_i}
\mathrm{(P_3)}~
\lefteqn{T_{i}^{*}(t_i)=\min_{\textbf{S}_i}
\frac{\lambda_i-t_i}{\lambda_i}\frac{\bar{r_i}}{f_i}+\frac{1}{\lambda_i}\sum\limits_{j=1}^{N}\tilde{\lambda_{i,j}}\Bigg(
\phi_{i,j}+\frac{1}{2}
}\nonumber\\
&\times&\frac{\theta_{i,j}\frac{\tilde{\lambda_{i,j}}^2}{t_i}+\tilde{\lambda_{i,j}}\alpha_{i,j}+\frac{\tilde{\lambda_{i,j}}}{t_i}
\beta_{i,j}+\gamma_{i,j}}
{1-\left(\phi_{i,j}\frac{\tilde{\lambda_{i,j}}^2}{t_i}+\tilde{\lambda_{i,j}}\delta_{i,j}+\frac{\tilde{\lambda_{i,j}}}{t_i} \mu_{i,j}+\nu_{i,j}\right)}
\Bigg)~~~~~
\end{eqnarray}
\begin{subequations}
\begin{numcases}{ \mbox{s.t.}}
\tilde{\lambda}_{i,j}\ge0, \forall j\in \mathbb{N},\\
\sum\limits_{j=1}^{N}\tilde{\lambda_{i,j}}=t_i,\label{}\\
\phi_{i,j}\frac{\tilde{\lambda_{i,j}}^2}{t_i}+\tilde{\lambda_{i,j}}\delta_{i,j}+\frac{\tilde{\lambda_{i,j}}}{t_i}
\mu_{i,j}+\nu_{i,j}-1 \le0,\label{}\forall j~~~~~~\\
\sum\limits_{j=1}^N\tilde{\lambda}_{i,j}\left(P_{i,j}\frac{\bar{A_{i,j}}}{R_{i,j}}+\frac{\bar{r_i}}{f_i}\eta_if_{i}^3\right)+P_s-\mathcal{E}_i- \mathfrak{C}_i\le0.~~~~~~~~\label{}
\end{numcases}
\end{subequations}
Next, find the optimal $t_i^*=\inf\limits_{t_i\in \left[0,\lambda_i\right]}T_{i}^{*}(t_i)$ through one-dimensional search.
\subsection{Game analysis}\label{Sec:Game analysis}
In the paper, we consider the computation tasks and the response time in average sense, i.e., depicted by the arriving rate and average response time, respectively. Therefore mixed strategy is appropriate for the game analysis.
To begin with, the mixed strategy NE of the formulated noncooperative game $\mathcal{G}$ is described in Definition \ref{Definition: mixed NE}.

Let $\Pi_i$ be the set of probability measures over the pure strategy (action) set $\mathcal{S}_i$. $\sigma_i\in \Pi_i$ is a mixed strategy of MD $i$, and $\sigma_{-i}$ denotes the set of mixed strategies of players except $i$.
\begin{definition}\label{Definition: mixed NE}
$\sigma^*=\left(\sigma_1^*,\cdots,\sigma_M^*\right)\in \Pi=\Pi_1\times\cdots\times\Pi_M$ ($\times$ is Cartesian Product) is a mixed strategy NE of $\mathcal{G}$ if $T_i\left(\sigma_i^*,\sigma_{-i}^*\right)\le T_i\left(\sigma_{i},\sigma_{-i}^*\right)$ for $\sigma\in\Pi$ and $i\in \mathbb{M}$.
\end{definition}

\begin{property}\label{property: mixed NE}
If $T_i\left(\sigma_i^*,\sigma_{-i}^*\right)\le T_i\left(\textbf{S}_i,\sigma_{-i}^*\right)$ for $\textbf{S}_i\in\mathcal{S}_i$ and $i\in \mathbb{M}$, $\sigma^*$ is a mixed strategy NE.
\end{property}
\begin{IEEEproof}
According to the von Neumann-Morgenstern expected payoff theory, we have $$T_i\left(\sigma_{i},\sigma_{-i}^*\right)=\int_{\mathcal{S}_i}T_i\left(\textbf{S}_i,\sigma_{-i}^*\right)d\sigma_{i}(\textbf{S}_i).$$ That is to say, only pure strategy deviations when determining whether a given profile is an NE. Hence, we have Property \ref{property: mixed NE}.
\end{IEEEproof}
Regarding the existence, we have the following Lemma.

\begin{lemma}
$\mathcal{G}$ has at least one mixed strategy NE.
\end{lemma}
\begin{IEEEproof}
The action set $\mathcal{S}_i$ is nonempty and compact, the payoff function $T_i$ is continuous. According to \cite{I. L. Glicksberg1952:Proceedings of the National Academy of Sciences}, the game has a mixed strategy NE.
\end{IEEEproof}

\section{Algorithm Design}\label{Sec:Algorithm design}
In this section, we turn our attention to solve the shrunken problem $\mathrm{(P_3)}$ at first. After that, we conceive an NE-orienting algorithm.

$\mathrm{(P_3)}$ is a fractional programming problem, and is not convex in general. The KKT conditions can be derived in Lemma \ref{Lemma:KKT Conditions for suboptimization}.
\begin{lemma}\label{Lemma:KKT Conditions for suboptimization}
For optimization problem $\mathrm{(P_3)}$, the KKT conditions can be described as
\begin{subequations}\label{KKT Conditions for suboptimization}
\begin{numcases}{}
(\ref{Partial for Suboptimization}), \forall j\label{}\\
\tilde{\lambda_{i,j}}\ge 0,\forall j\\
\epsilon_j\ge0,\forall j\\
\epsilon_j\tilde{\lambda_{i,j}}=0,\forall j\\
  \sum\limits_{j=1}^N\tilde{\lambda}_{i,j}-t_i=0,\label{}\\
\frac{\phi_{i,j}}{t_i}\tilde{\lambda_{i,j}}^2+\left(\delta_{i,j}+\frac{\mu_{i,j}}{t_i}\right)\tilde{\lambda_{i,j}}
+\nu_{i,j}-1 \le0, \forall j~~\label{}\\
  \varpi_j\ge0,\forall j\\
  \varpi_j\Big[\frac{\phi_{i,j}}{t_i}\tilde{\lambda_{i,j}}^2+\left(\delta_{i,j}+\frac{\mu_{i,j}}{t_i}\right)\tilde{\lambda_{i,j}}
+\nu_{i,j}-1\Big]=0,\forall j~~~~~~~~\\
 \sum\limits_{j=1}^N\tilde{\lambda}_{i,j}\left(P_{i,j}\frac{\bar{A_{i,j}}}{R_{i,j}}+\frac{\bar{r_i}}{f_i}\eta_if_{i}^3\right)+P_s-\mathcal{E}_i- \mathfrak{C}_i\le0,~~~ \\
  \rho\ge 0,\\
  \rho\Big[\sum\limits_{j=1}^N\tilde{\lambda}_{i,j}\left(P_{i,j}\frac{\bar{A_{i,j}}}{R_{i,j}}+\frac{\bar{r_i}}{f_i}\eta_if_{i}^3\right)+P_s-\mathcal{E}_i- \mathfrak{C}_i\Big]=0, \nonumber\\
\end{numcases}
\end{subequations}
where $\epsilon_j$, $\varpi_j$, and $\rho$ are multipliers,
$X_{i,j}=\phi_{i,j}-\epsilon_j+\chi+\rho\left(P_{i,j}\frac{\bar{A_{i,j}}}{R_{i,j}}+\frac{\bar{r_i}}{f_i}\eta_if_{i}^3\right)+\varpi_j(\frac{\mu_{i,j}}{t_i}+\delta_{i,j})$.
\end{lemma}

\begin{figure*}[t]
\begin{eqnarray}\label{Partial for Suboptimization}
&&4\varpi_j\frac{\phi_{i,j}^3}{t_i}\tilde{\lambda_{i,j}}^5+
\Big[2X_{i,j}\phi_{i,j}^2+8\varpi_j\frac{\phi_{i,j}^2}{t_i}(\mu_{i,j}+\delta_{i,j}t_i)-5\theta_{i,j}\phi_{i,j}\Big]\tilde{\lambda_{i,j}}^4
\nonumber\\
&+&\Big\{4X_{i,j}\phi_{i,j}(\mu_{i,j}+\delta_{i,j}t_i)+4\varpi_j\frac{\phi_{i,j}}{t_i}\big[2\phi_{i,j}(t_i-\mu_{i,j}t_i)+(\mu_{i,j}+\delta_{i,j}t_i)^2\big]-4\theta_{i,j}(\mu_{i,j}
+\delta_{i,j}t_i)-4\phi_{i,j}(\beta_{i,j}+\alpha_{i,j}t_i)\Big\}\tilde{\lambda_{i,j}}^3\nonumber\\
&+&\Big\{2\big[2\phi_{i,j}(t_i-\mu_{i,j}t_i)+(\mu_{i,j}+\delta_{i,j}t_i)^2\big]X_{i,j}+8\varpi_j\frac{\phi_{i,j}}{t_i}(\beta_{i,j}+\alpha_{i,j}t_i)(t_i-\nu_{i,j}t_i)+3\big[\theta_{i,j}(t_i-\nu_{i,j}t_i)-\phi_{i,j}\gamma_{i,j}t_i-(\beta_{i,j}\nonumber\\
&+&\alpha_{i,j}t_i)(\mu_{i,j}+\delta_{i,j}t_i)\big]\Big\}\tilde{\lambda_{i,j}}^2+\Big[4X_{i,j}(\mu_{i,j}+\delta_{i,j}t_i)(t_i-\nu_{i,j}t_i)+4\varpi_j\frac{\phi_{i,j}}{t_i}(t_i-\nu_{i,j}t_i)^2+2(\beta_{i,j}+\alpha_{i,j}t_i)(t_i-\nu_{i,j}t_i)\nonumber\\
&-&2\gamma_{i,j}t_i(\mu_{i,j}+\delta_{i,j}t_i)\Big]\tilde{\lambda_{i,j}}+2(t_i-\nu_{i,j}t_i)^2X_{i,j}+\gamma_{i,j}t_i(t_i-\nu_{i,j}t_i)\nonumber\\
&=&0,
\end{eqnarray}
\end{figure*}

\begin{IEEEproof}
See Appendix \ref{Appendix:suboptimization KKT proof}.
\end{IEEEproof}
Compared with (\ref{KKT Condition}), (\ref{KKT Conditions for suboptimization}) has been greatly reduced. The variables $\{\tilde{\lambda_{i,1}},\cdots,\tilde{\lambda_{i,N}}\}$ have been decoupled. Each variable can be obtained independently by solving a quintic equation with constant coefficients.
Generally, KKT conditions give necessary but NOT sufficient solutions. However, (\ref{KKT Conditions for suboptimization}) has finite solutions. A quintic equation has five solutions, then we have at most $5^N$ combinations. By exhausting these combinations, we can get the optimal one.\footnote{ The time efficiency is not high especially when $N$ is large.}

We design Algorithm 1 to attain the best response of an MD when other MDs' offloading policies are fixed. Mathematically, Algorithm 1 is plotted to get the solutions of optimization problem $\mathrm(P_2)$. The proposed 2-stage strategy is utilized. In Algorithm 1, we depict a golden section search method for the one-dimensional search in handling $t_i^*=\inf\limits_{t_i\in \left[0,\lambda_i\right]}T_{i}^{*}(t_i)$. For each $t_i$, $T_{i}^{*}(t_i)$ is obtained by acquiring the solution of the KKT conditions (\ref{KKT Conditions for suboptimization}). Finally, we get the optimal offloading policy and the average response time.
\begin{table}[]
 \centering
 \begin{tabular}{lcl}
  \toprule
  \textbf{Algorithm 1: Optimal offloading algorithm for MD $i$}\\
  ~~~~~~\textbf{given other MDs' offloading policies}\\
  \midrule
 Step 1:  $k=0$, initialize $a=0$, $b=\lambda_i$,
 $a_1=b-0.618*(b-a)$, \\
 $a_2=a+0.618*(b-a)$, $f_1=T_{i}^{*}(a_1)$,
 $f_2=T_{i}^{*}(a_2)$.\\
 Step 2:  if $f_1>f_2$, \\
 $a=a_1$, $a_1=a_2$, $f_1=f_2$, $a_2=a+0.618*(b-a)$, $f_2=T_{i}^{*}(a_2)$.\\
else\\
$b=a_2$, $a_2=a_1$, $f_2=f_1$, $a_1=b-0.618*(b-a)$, $f_1=T_{i}^{*}(a_1)$.\\
 Step 3: $k=k+1$, go to Step 2 until convergence. \\
 Step 4: $t^*=\frac{a+b}{2}$. The optimal policy is the solution of (\ref{KKT Conditions for suboptimization}) given $t^*$.\\
  \bottomrule
 \end{tabular}
\end{table}

In Algorithm 2, each MD has a feasible initial offloading policy, $\left\{\mathbf{S}_i^{(0)}\right\}_{i\in\mathbb{M}}$. In the $k$-th iteration, MD $i$ derives its optimal offloading policy, $\mathbf{S}_i^{(k+1)}$, by applying Algorithm 1, under the circumstances that other MDs employ offloading policies in $(k-1)$-th iterations, $\mathbf{S}_{-i}^{(k)}$. All $M$ MDs complete the renewals in the iteration. The execution of the iterative algorithm terminates when satisfying a stoping criterion. Denote the outcome of Algorithm 2 as $\mathbf{S}_i^*=(\tilde{\lambda}_{i,1}^{*},\cdots,\tilde{\lambda}_{i,N}^{*})$. We have the following mixed strategy NE solution.
 When a task arrives at MD $i$, it is locally processed with probability $\frac{\lambda_i-\sum\limits_{j=1}^{N}\tilde{\lambda}_{i,j}^*}{\lambda_i}$ and offloaded to MEC-s $j$ with probability
$\frac{\tilde{\lambda}_{i,j}^*}{\lambda_i}$.

\begin{table}[]
 \centering
 \begin{tabular}{lcl}
  \toprule
  \textbf{Algorithm 2: Iterated distributive NE-orienting algorithm}\\
  \midrule
 Step 1:  $k=0$, initialize feasible policy for $M$ MDs, $\left\{\mathbf{S}_i^{(0)}\right\}_{i\in\mathbb{M}}$.\\
 Step 2: For every $i \in \mathbb{M}$, compute the offloading policy \\
 ~~$\mathbf{S}_i^{(k+1)}$ by applying Algorithm 1, given $\mathbf{S}_{-i}^{(k)}$. \\
 Step 3: $k=k+1$, go to Step 2 until convergence.\\
  \bottomrule
 \end{tabular}
\end{table}

\section{Numerical results}\label{Sec:Numerical results}
In this section, simulations are performed to verify and exhibit the effectiveness and efficiency of the proposed algorithm. We first explore the convergence performance, and then the average response time regarding parameters such as the computation capability of MEC-s, data transmission rate, and computation task generation rate are investigated, respectively.

In the simulations, we discuss the 2-MD and 2-MEC-s scenario. The parameters are listed in Table \ref{Parameters of sim} unless specified otherwise.
 \begin{table}[]
\caption{\label{Parameters of sim}}
\centering
\begin{tabular}{cl|cl}
 \hline
   \hline
  $\lambda_1$ & $5$ & $\lambda_2$& $4$\\
 $f_1$ & $2$ & $f_2$& $2$\\
  $F_1$ & $20$ & $F_2$& $18$\\
  $\bar{r}_1$&$1.5$ & $\bar{r}_2$&$1.1$\\
  $\bar{r}_1^2$&0.7&$\bar{r}_2^2$&0.9\\
  $\bar{A}_{1,1}$&1&$\bar{A}_{1,2}$&1\\
  $\bar{A}_{1,1}^2$&0.6& $\bar{A}_{1,2}^2$&0.4\\
  $\bar{A}_{2,1}$&1& $\bar{A}_{2,2}$&1\\
  $\bar{A}_{2,1}^2$&0.5& $\bar{A}_{2,2}^2$&0.5\\
  $\eta_1$&0.55& $\eta_2$&0.6\\
  $\mathcal{E}_1$&12& $\mathcal{E}_2$&11\\
  $\mathfrak{C}_1$&80& $\mathfrak{C}_2$&85\\
  $R_{1,1}$&7& $R_{1,2}$&5\\
  $R_{2,1}$&7& $R_{2,2}$&6\\
  $P_s$&0.3&$N_0$&0.1\\
  $B$&10&&\\
\hline
\end{tabular}
\end{table}
 We consider Rayleigh fading environment, the power gains of the wireless channel, $\{|h_{i,j}|^2\}_{i=1,2}^{j=1,2}$, are exponentially distributed with mean $0.3, 0.2, 0.25, 0.25$, respectively. $P_{i,j}$ ($i=1,2,j=1,2$) can be computed according to (\ref{Rate power relation}).
Observe that $\phi_{i,j}\frac{\tilde{\lambda_{i,j}}^2}{t_i}+\tilde{\lambda_{i,j}}\delta_{i,j}+\frac{\tilde{\lambda_{i,j}}}{t_i}
\mu_{i,j}+\nu_{i,j}-1=0$ should not reach for the optimal solutions. Otherwise, the objective function value is infinite. Therefore, the corresponding Lagrange multiplier $\varpi_j=0$. When (\ref{KKT Conditions for suboptimization}) has no solutions, arbitrary allocation policy that satisfies (\ref{ShrinkedKKT}) is adopted. Once (\ref{ShrinkedKKT}) has on solutions, it means that the power budget is NOT sufficient for offloading $t$ tasks under the setting rate.
\begin{subequations}\label{ShrinkedKKT}
\begin{numcases}{}
  \sum\limits_{j=1}^N\tilde{\lambda}_{i,j}-t_i=0,\label{}\\
  \sum\limits_{j=1}^N\tilde{\lambda}_{i,j}\left(P_{i,j}\frac{\bar{A_{i,j}}}{R_{i,j}}+\frac{\bar{r_i}}{f_i}\eta_if_{i}^3\right)+P_s-\mathcal{E}_i- \mathfrak{C}_i\le0.~~~~~
\end{numcases}
\end{subequations}

\begin{figure}[]
\centering
\subfigure[$|h_{1,1}|^2=0.1375$, $|h_{1,2}|^2=0.4655$, $|h_{2,1}|^2=0.3196$, $|h_{2,2}|^2=0.1509$.]{\includegraphics[angle=0,width=3.0in]{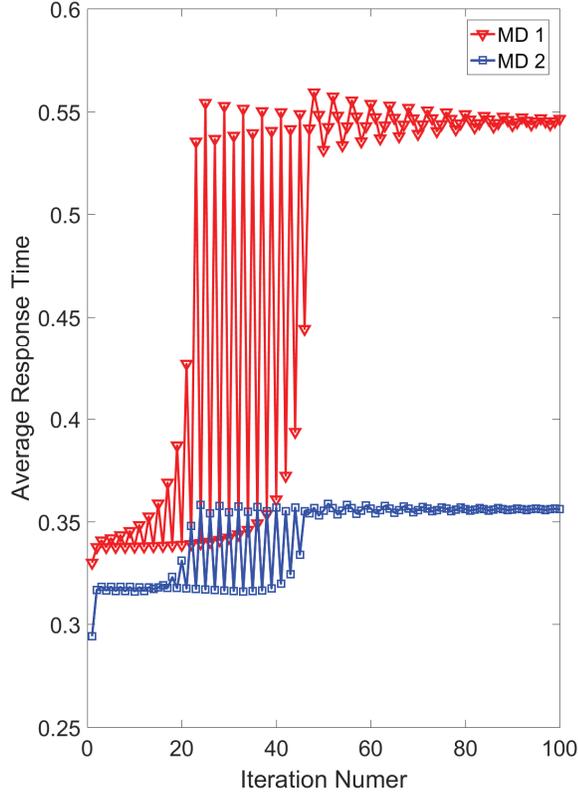}}
\subfigure[$|h_{1,1}|^2=0.0615$, $|h_{1,2}|^2=0.0198$, $|h_{2,1}|^2=0.5159$, $|h_{2,2}|^2=0.0227$.]{\includegraphics[angle=0,width=3.0in]{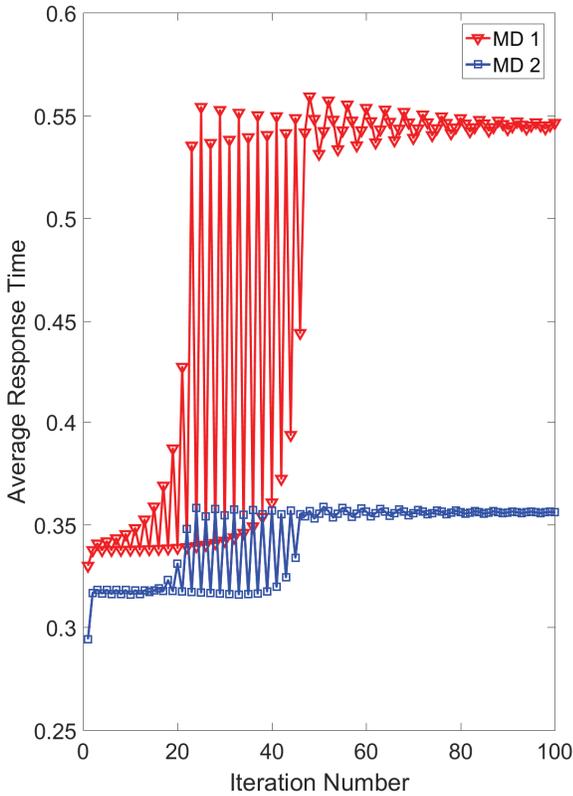}}
\caption{Convergence performance of Algorithm 2}
\label{Sim:Convergence}
\end{figure}

Fig. \ref{Sim:Convergence} demonstrates the convergence performance of the proposed algorithm. Two wireless channel configurations are given. From both figures, we can see that Algorithm 2 converges since 90-th iteration. In the figures, MD 2 has better average response time performance. It is because that MD 1 has higher computation task generation rate.
Another observation is that the channel configuration has few influences on the convergence performance. This can be explained as follows: In the simulation settings, we have fixed transmission rate $R_{i,j}$. The channel power gain $|h_{i,j}|^2$ is related to the transmitting power of the mobile devices $P_{i,j}$. $P_{i,j}$ is related to the constraint (\ref{power constraint}). When (\ref{power constraint}) is lose, the channel power gain $|h_{i,j}|^2$ has no effect on the formulated problem. Therefore, the two wireless channel configurations in the figure have similar convergence performance.

\begin{figure*}[]
\centering
\subfigure[$F_1=20*c$.]{\includegraphics[angle=0,width=3.0in]{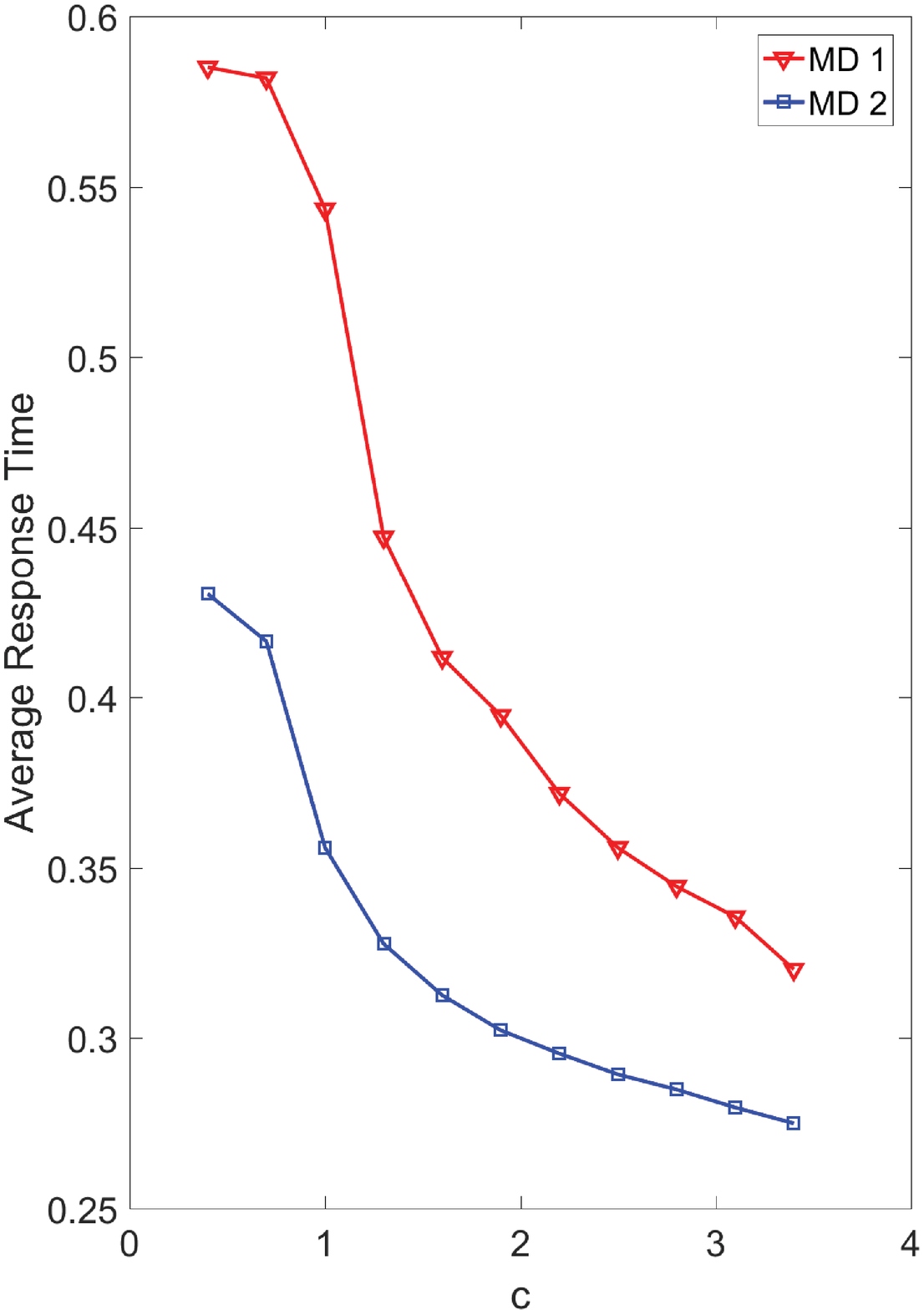}\label{Fig:F1}}
\subfigure[$F_2=18*c$.]{\includegraphics[angle=0,width=3.0in]{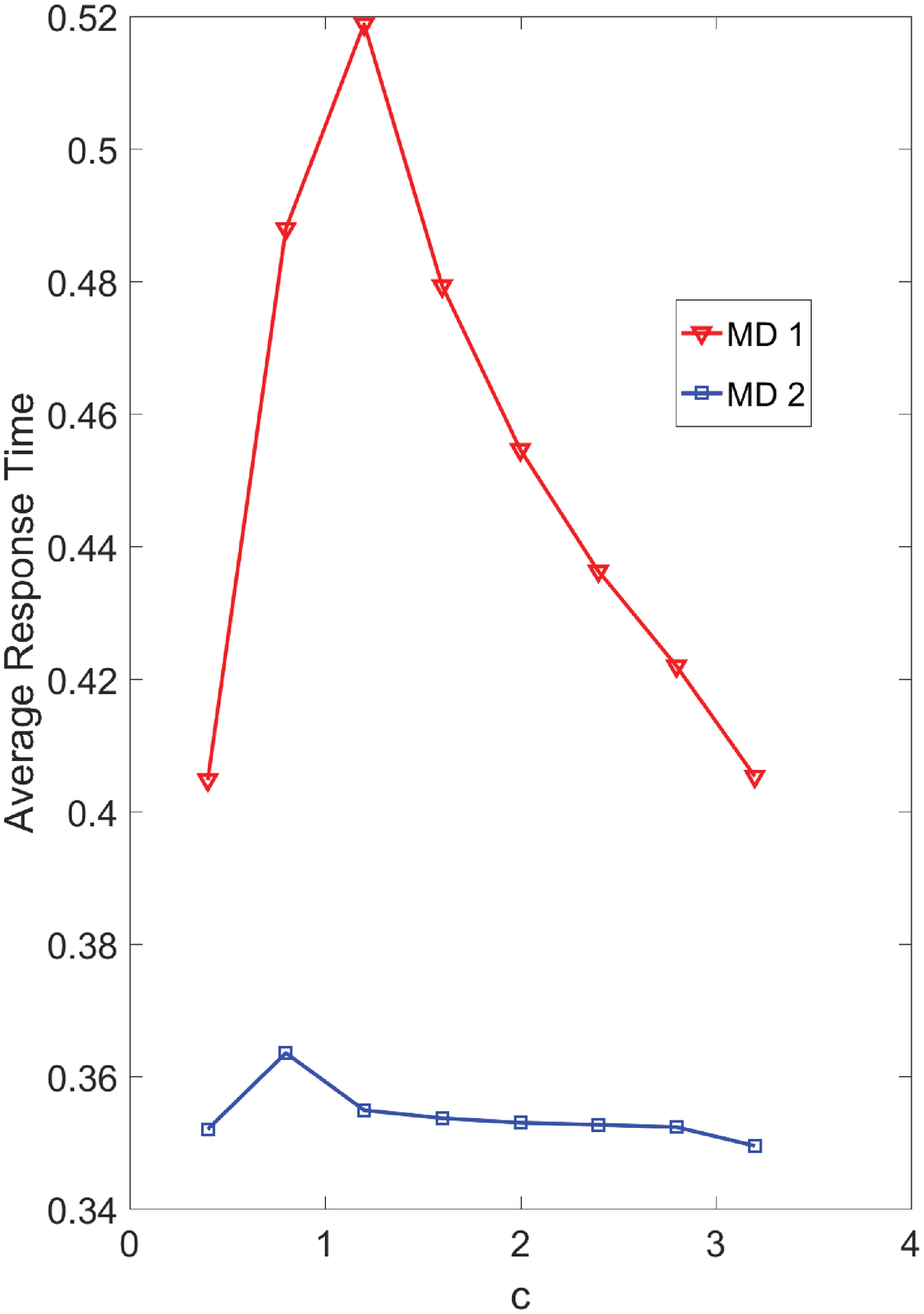}\label{Fig:F2}}
\subfigure[$F_1=20*c$, $F_2=18*c$.]{\includegraphics[angle=0,width=3.0in]{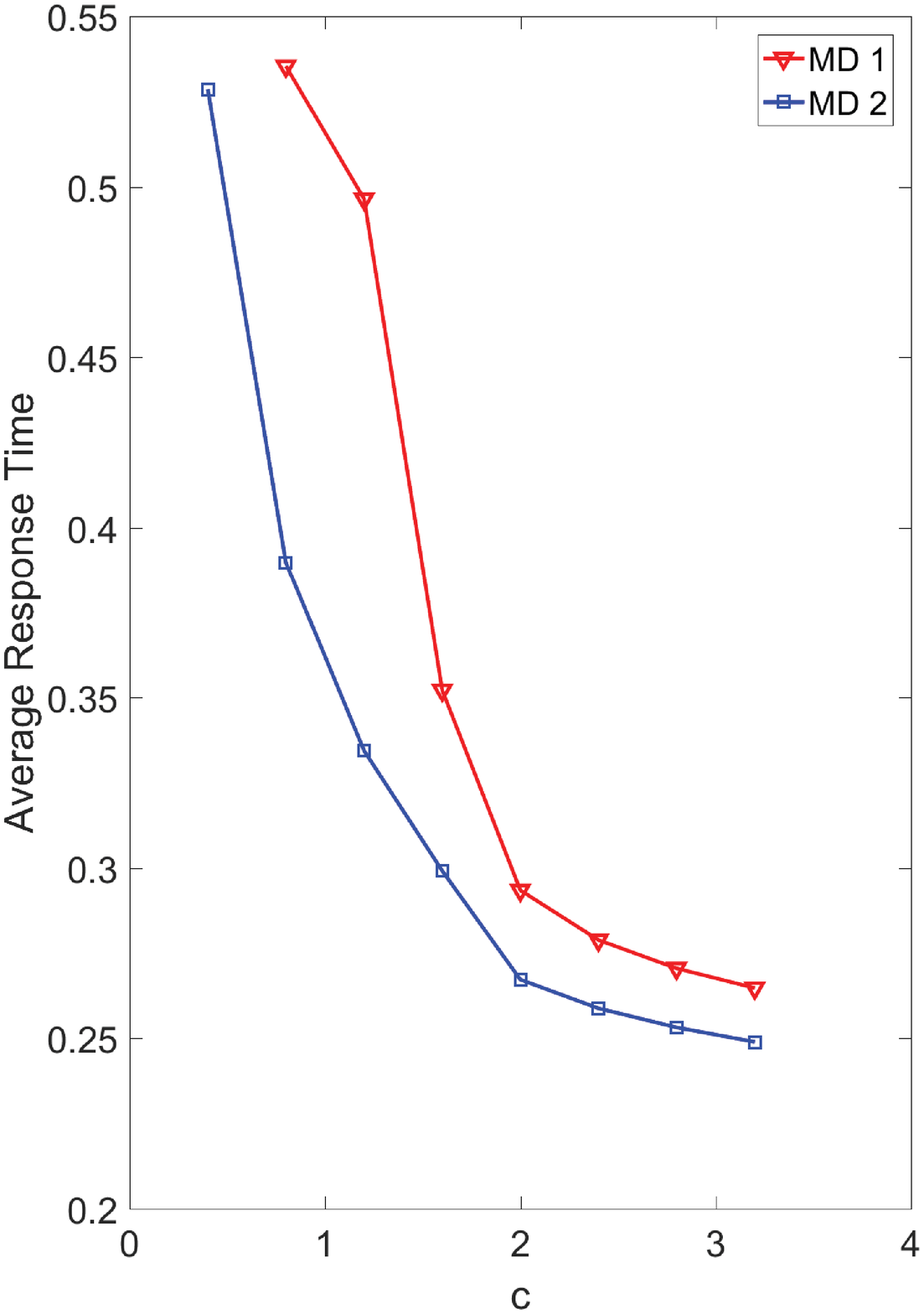}\label{Fig:F3}}
\caption{Average response time vs. the computation speed of MEC-s }
\label{Sim:regarding Computation speed of MEC-s}
\end{figure*}

Fig. \ref{Sim:regarding Computation speed of MEC-s} plots the average response time with respect to the computation capability of MEC-s. $F_1$, $F_2$, and $F_1$ together with $F_2$ vary with the coefficient $c$ in Fig. \ref{Fig:F1}, Fig. \ref{Fig:F2}, and Fig. \ref{Fig:F3}, respectively.  In Fig. \ref{Fig:F1} and Fig. \ref{Fig:F3}, the average repones time decreases with the increment of   $c$, harshly at first and smoothly later.
This is because that with better computation capability of MEC-s $1$ and MEC-s $2$, the processing time and the waiting time at MEC-s become shorter (the data transmission time remains static) under some computation offloading strategy. Then the average response time diminishes. At the beginning, the processing time and the waiting time dominate in the average response time, and the drop is apparent. With the decrement of the processing and waiting time, they are not dominant after a value of $c$. Then the decrement of the average response time becomes moderate. In Fig. \ref{Fig:F2}, we can observe that the average response time increases in a short range ($c=0.4, 0.8, 1.2$) at first and then decreases. The reasons are as follows. The offloading strategy alters for $c=0.4, 0.8, 1.2$. Under different allocation strategies, the average response time have a short range increment ($c=0.4, 0.8, 1.2$ for MD $1$ and $c=0.4, 0.8$ for MD $2$). When the offloading strategy remains static, the average response time decreases with the increment of $F_2$. For MD $1$, the increment is harsh, and for MD $2$, moderate. This shows that more computation tasks of MD 1 are offloaded to MEC-s 2. As for MD 2, more tasks are processed at MEC-s 1 than MEC-s 2.

\begin{figure}[]
\centering
\subfigure[$R_{1,1}=7*c$.]{\includegraphics[angle=0,width=3.0in]{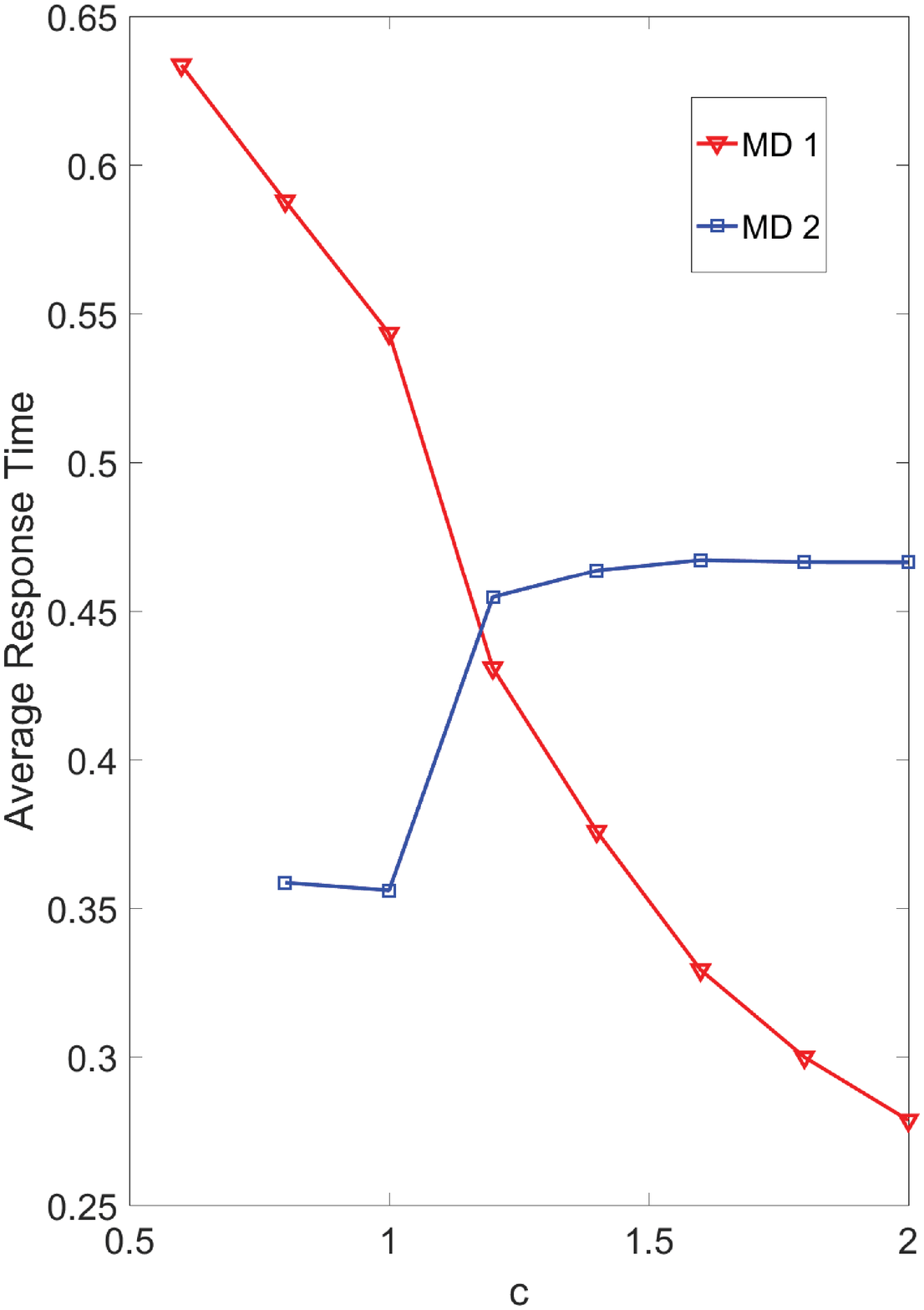}\label{Fig:R11}}
\subfigure[$R_{2,2}=6*c$.]{\includegraphics[angle=0,width=3.0in]{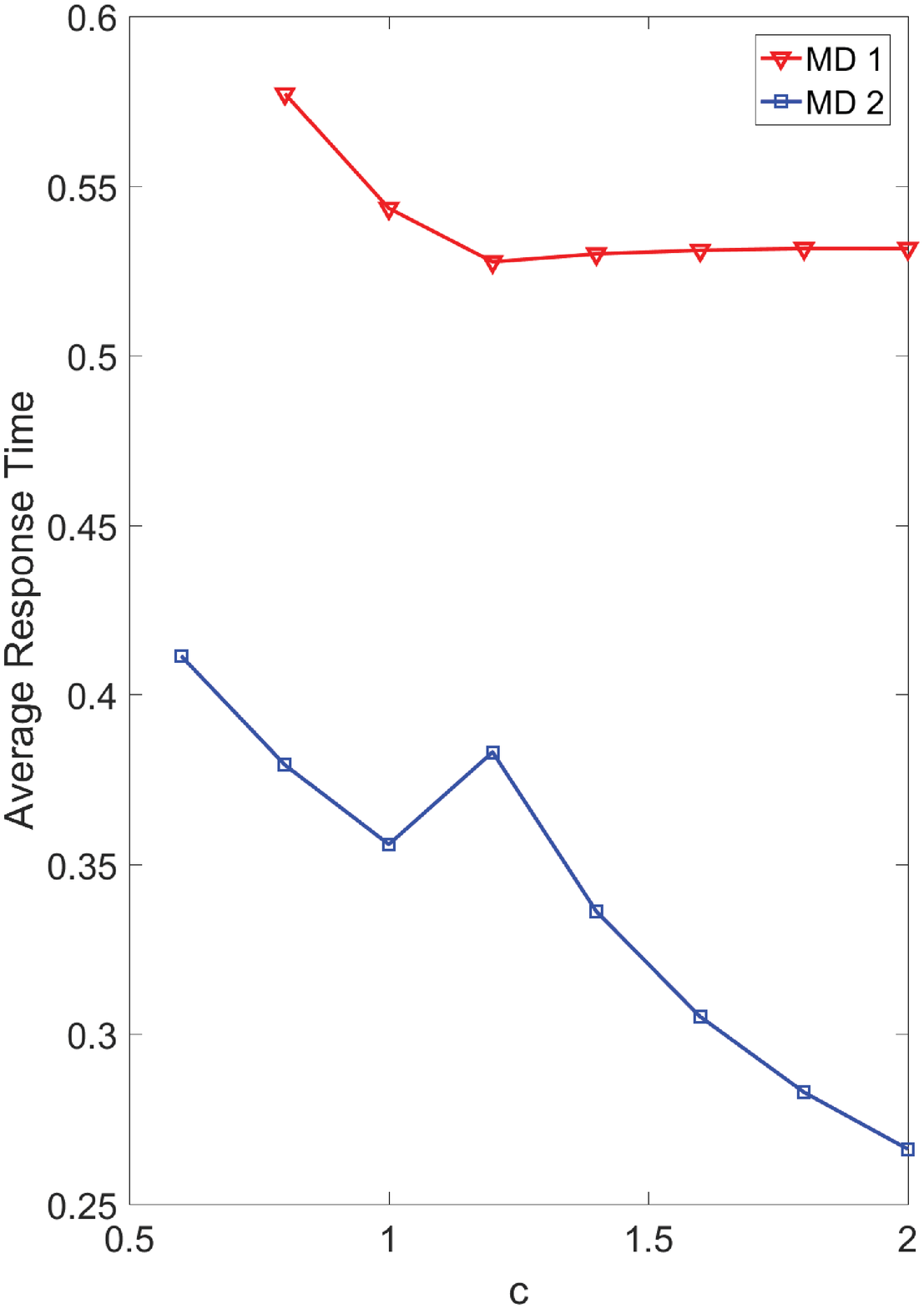}\label{Fig:R22}}

\caption{Average response time vs. data transmission rate}
\label{Sim:regarding rate}
\end{figure}

Fig. \ref{Sim:regarding rate} draws the average response time regarding the data transmission rate. $R_{1,1}$ and $R_{2,2}$ alter with the coefficient $c$ in Fig. \ref{Fig:R11} and Fig. \ref{Fig:R22}, respectively. In Fig. \ref{Fig:R11}, we can find that with the increment of $R_{1,1}$, the response time of MD $1$ lessens apparently in the commence and then moderately. Meanwhile, the average response time of MD 2 has an increment at first and remains constant then. The reasons are as follows. When $R_{1,1}$ increases, the data transmission time of MD 1 decreases. At the same time, the portion of the transmission time in the average response time (data transmission time plus processing time plus waiting time) becomes smaller. Then the decrement is drastic initially and moderate then. Regarding MD 2, the offloading policy changes at first, and the average response time increases. When the offloading policy remains static, $R_{1,1}$ has few effects on its average response time. In Fig. \ref{Fig:R22}, the average response time of MD 2 decreases with the increment of $R_{2,2}$ for similar reasons as in Fig. \ref{Fig:R11}. The difference is that there is a increment from $c=1$ to $c=1.2$ since the offloading policy alteration. The average response time of MD 1 decreases firstly since the offloading policy shift, and remains static as $R_{2,2}$ has few effects on its average performance under fixed offloading policy.

\begin{figure}[]
\centering
\subfigure[$\lambda_1=5*c$, $\bar{r}_1=1.5*c$, $\bar{r}^2_1=0.7*c^2$, $\bar{A}_{1,1}=1*c$, $\bar{A}_{1,2}=1*c$, $\bar{A}_{1,1}^2=0.6*c^2$, and $\bar{A}_{1,2}^2=0.4*c^2$.]{\includegraphics[angle=0,width=2.9in]{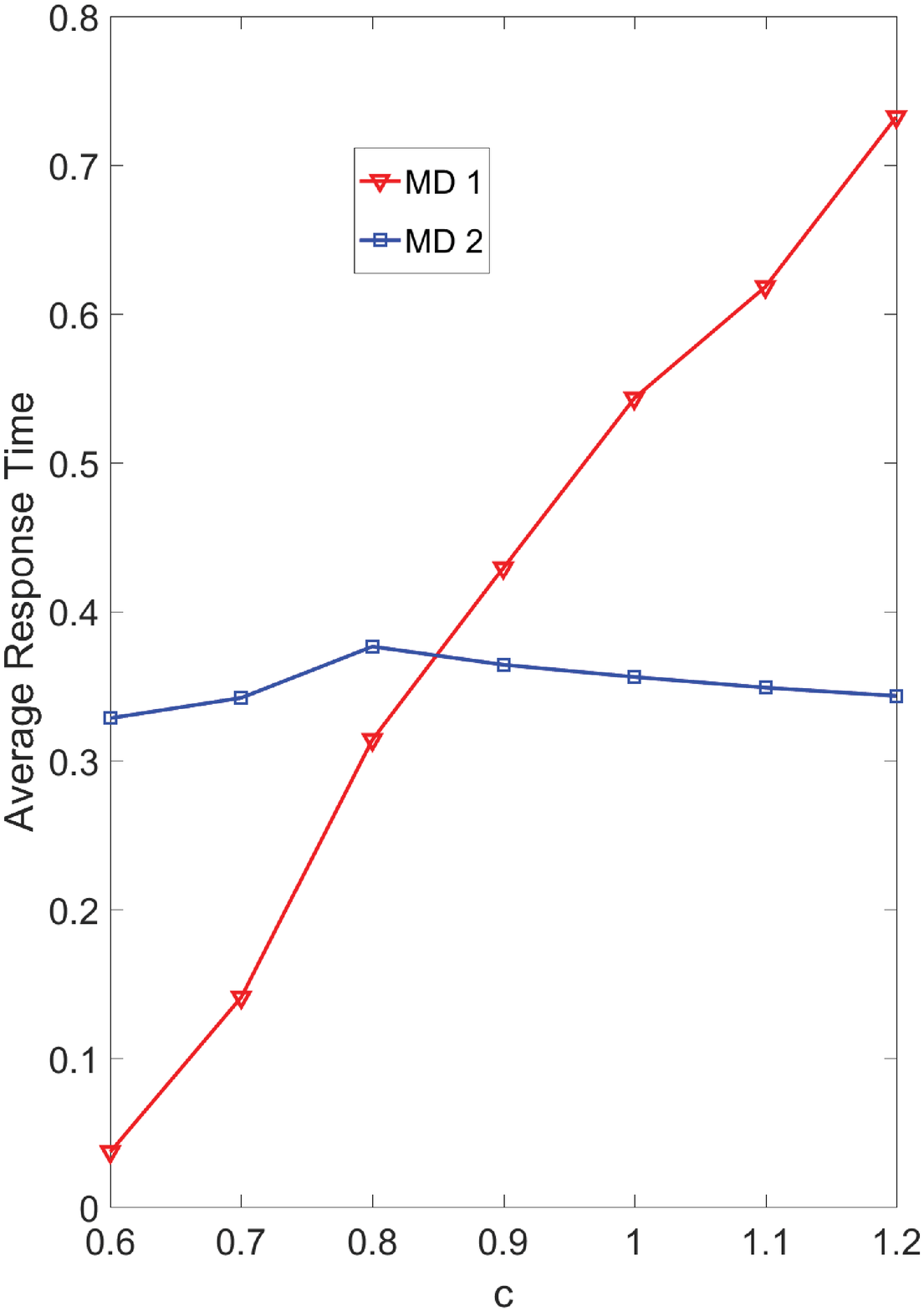}\label{Fig:Lamda1}}
\subfigure[$\lambda_2=4*c$, $\bar{r}_2=1.1*c$, $\bar{r}^2_2=0.9*c^2$, $\bar{A}_{2,1}=1*c$, $\bar{A}_{2,2}=1*c$, $\bar{A}_{2,1}^2=0.5*c^2$, and $\bar{A}_{2,2}^2=0.5*c^2$.]{\includegraphics[angle=0,width=2.9in]{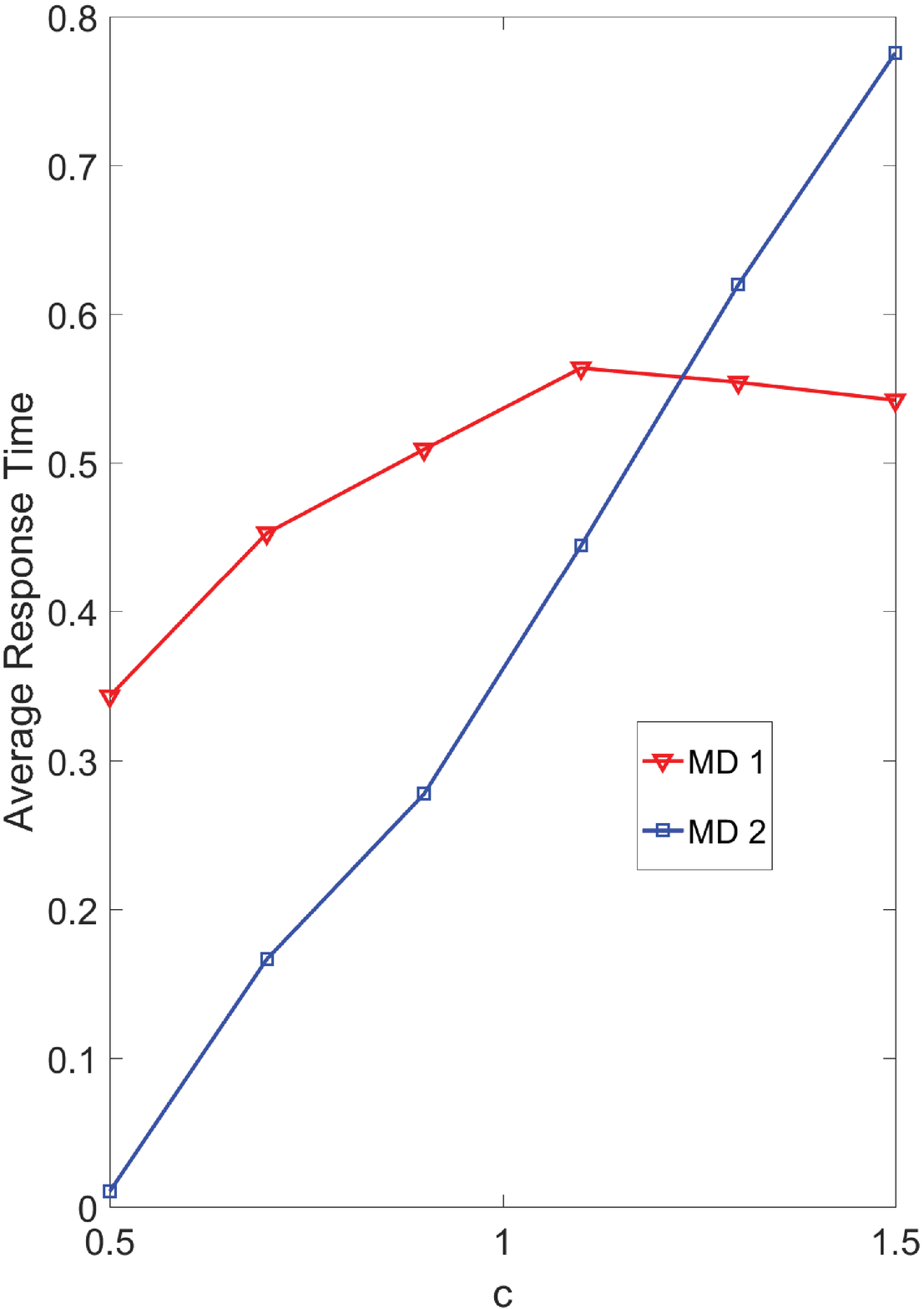}\label{Fig:Lamda2}}

\caption{Average response time vs. computation task generation rate}
\label{Sim:regarding task rate}
\end{figure}

Fig. \ref{Sim:regarding task rate} shows the average response time when the computation task generation rate modifies.
From Fig. \ref{Fig:Lamda1} and Fig. \ref{Fig:Lamda2}, we can see that when the computation task generation rate of MD 1 (2) increases, its average response time increases apparently. It is because the processing time, the waiting time, and the transmission time increase with the augment of task generate rate. For MD 2 (1), its average response time expands mildly at the beginning and modifies slightly then. This can be explained as follows. When MD 1 (2) has more tasks for offloading, the waiting time of MD 2 (1) increases under fixed offloading strategy. Thus the average response time of MD 2 (1) increases at first. When the average response time is larger than a value, MD 2 (1) changes its offloading policy to alleviate the influence, and the response time changes slightly.

\section{Conclusion}\label{Sec:conclusion}
In the paper, we discuss the computation offloading of multi-MD to multi-MEC-s over wireless interference channels in energy harvesting aided heterogeneous MEC networks. Considering the stochastic features of computation task generation, wireless channel, harvested energy, queueing time, the interference among MDs, the transmission power budget, and the average response time minimization aim of each MD,
we establish a noncooperative computation offloading game framework. Via the problem reconstruction, the 2-step decomposition, and the solution of KKT conditions, the optimization of one MD is settled. Thereby, we present an iterated distributive NE-orienting algorithm. The convergence performance and effect of parameters are investigated in the simulations. Future topics include considering the transmission rate allocation together with computation offloading, the time efficiency improvement, etc.


%

\appendices
\section{Proof of Lemma \ref{Lemma:KKT Condition}}\label{Appendix:KKT proof}
The KKT conditions of $\mathrm{(P_1)}$ are written as
\begin{subequations}
\begin{numcases}{}
\nabla T_i-\sum\limits_{j}\epsilon_j\nabla (\tilde{\lambda_{i,j}})+\chi\nabla(\tilde{\lambda}_{i}-\lambda_i)+\sum\limits_{j}\varpi_j\nabla(\zeta_j-1)\nonumber\\
+\rho\nabla(P_{i}-\mathcal{E}_i- \mathfrak{C}_i)=0,\label{}\\
\tilde{\lambda_{i,j}}\ge 0,\forall j\\
\epsilon_j\ge0,\forall j\\
\epsilon_j\tilde{\lambda_{i,j}}=0,\forall j\\
  \tilde{\lambda}_{i}-\lambda_i\le0,\label{}\\
  \chi\ge 0,\\
  \chi (\tilde{\lambda}_{i}-\lambda_i)=0,\\
  \zeta_j-1\le0,\label{} \forall j~~~~~~~~\\
  \varpi_j\ge0,\forall j\\
  \varpi_j\Big(\zeta_j-1\Big)=0,\forall j~~~~~~\\
  P_{i}-\mathcal{E}_i- \mathfrak{C}_i\le 0,\\
  \rho\ge 0, \\
  \rho(P_{i}-\mathcal{E}_i- \mathfrak{C}_i)=0
 \label{}
\end{numcases}
\end{subequations}
$T_i$ can be reexpressed as
\begin{eqnarray}
T_i=\frac{\lambda_i-\sum\limits_{l=1}^{N}\tilde{\lambda_{i,l}}}{\lambda_i}\frac{\bar{r_i}}{f_i}+\frac{1}{\lambda_i}\sum\limits_{j=1}^{N}\tilde{\lambda_{i,j}}\left(
\phi_{i,j}
+\frac{1}{2}H_{i,j}\right),
\end{eqnarray}
where $H_{i,j}$ is given by (\ref{H_ij}).
\begin{figure*}
\begin{eqnarray}
\lefteqn{
H_{i,j}=\frac{\theta_{i,j}\frac{\tilde{\lambda_{i,j}}^2}{\sum\limits_{l=1}^{N}\tilde{\lambda_{i,l}}}+\tilde{\lambda_{i,j}}\alpha_{i,j}+\frac{\tilde{\lambda_{i,j}}}{\sum\limits_{l=1}^{N}\tilde{\lambda_{i,l}}}
\beta_{i,j}+\gamma_{i,j}}
{1-\left(\phi_{i,j}\frac{\tilde{\lambda_{i,j}}^2}{\sum\limits_{l=1}^{N}\tilde{\lambda_{i,l}}}+\tilde{\lambda_{i,j}}\delta_{i,j}+\frac{\tilde{\lambda_{i,j}}}{\sum\limits_{l=1}^{N}\tilde{\lambda_{i,l}}} \mu_{i,j}+\nu_{i,j}\right)}
} \nonumber\\
&=&\frac{\theta_{i,j}\tilde{\lambda_{i,j}}^2+\tilde{\lambda_{i,j}}\alpha_{i,j}\left(\sum\limits_{l=1}^{N}\tilde{\lambda_{i,l}}\right)+\tilde{\lambda_{i,j}}
\beta_{i,j}+\gamma_{i,j}\left(\sum\limits_{l=1}^{N}
\tilde{\lambda_{i,l}}\right)
}
{\sum\limits_{l=1}^{N}\tilde{\lambda_{i,l}}-\left(\phi_{i,j}\tilde{\lambda_{i,j}}^2+\tilde{\lambda_{i,j}}\delta_{i,j}\left(\sum\limits_{l=1}^{N}\tilde{\lambda_{i,l}}\right)+\tilde{\lambda_{i,j}}
\mu_{i,j}+\nu_{i,j}\left(\sum\limits_{l=1}^{N}
\tilde{\lambda_{i,l}}\right)\right)}\nonumber\\
&=&\frac{\theta_{i,j}\tilde{\lambda_{i,j}}^2+\tilde{\lambda_{i,j}}\alpha_{i,j}\left(\tilde{\lambda_{i,j}}+\sum\limits_{k\ne j,k\in \mathbb{N}}\tilde{\lambda_{i,k}}\right)+\tilde{\lambda_{i,j}}
\beta_{i,j}+\gamma_{i,j}\left(\tilde{\lambda_{i,j}}+\sum\limits_{k\ne j,k\in \mathbb{N}}\tilde{\lambda_{i,k}}\right)
}
{\sum\limits_{l=1}^{N}\tilde{\lambda_{i,l}}-\left(\phi_{i,j}\tilde{\lambda_{i,j}}^2+\tilde{\lambda_{i,j}}\delta_{i,j}\left(\tilde{\lambda_{i,j}}+\sum\limits_{k\ne j,k\in \mathbb{N}}\tilde{\lambda_{i,k}}\right)+\tilde{\lambda_{i,j}}
\mu_{i,j}+\nu_{i,j}\left(\tilde{\lambda_{i,j}}+\sum\limits_{k\ne j,k\in \mathbb{N}}\tilde{\lambda_{i,k}}\right)\right)}\nonumber\\
&=&\frac{\left(\theta_{i,j}+\alpha_{i,j}\right)\tilde{\lambda_{i,j}}^2+\left[\alpha_{i,j}\left(\sum\limits_{k\ne j,k\in \mathbb{N}}\tilde{\lambda_{i,k}}\right)+\gamma_{i,j}+\beta_{i,j}\right]\tilde{\lambda_{i,j}}+\gamma_{i,j}\left(\sum\limits_{k\ne j,k\in \mathbb{N}}\tilde{\lambda_{i,k}}\right)}
{\sum\limits_{l=1}^{N}\tilde{\lambda_{i,l}}-\left(\phi_{i,j}+\delta_{i,j}\right)\tilde{\lambda_{i,j}}^2-\left[\delta_{i,j}\left(\sum\limits_{k\ne j,k\in \mathbb{N}}\tilde{\lambda_{i,k}}\right)+\nu_{i,j}+\mu_{i,j}\right]\tilde{\lambda_{i,j}}-\nu_{i,j}\left(\sum\limits_{k\ne j,k\in \mathbb{N}}\tilde{\lambda_{i,k}}\right)}\label{H_ij}
\end{eqnarray}
\end{figure*}
Consequently,
\begin{eqnarray}\label{Partial Ti}
\frac{\partial{T_i}}{\partial{\tilde{\lambda_{i,j}}}}=-\frac{1}{\lambda_{i}f_i}+\frac{1}{\lambda_i}\left(
\phi_{i,j}
+\frac{1}{2}H_{i,j}
+\frac{1}{2}\tilde{\lambda_{i,j}}\frac{\partial{H_{i,j}}}{\partial{\tilde{\lambda_{i,j}}}}\right),
\end{eqnarray}
where $\frac{\partial{H_{i,j}}}{\partial{\tilde{\lambda_{i,j}}}}$ is given by (\ref{partial H_ij}).
\begin{figure*}
\begin{eqnarray}
\lefteqn{
\frac{\partial{H_{i,j}}}{\partial{\tilde{\lambda_{i,j}}}}=\frac{1-2\left(\phi_{i,j}+\delta_{i,j}\right)\tilde{\lambda_{i,j}}-\left[\delta_{i,j}\left(\sum\limits_{k\ne j,k\in \mathbb{N}}\tilde{\lambda_{i,k}}\right)+\nu_{i,j}+\mu_{i,j}\right]}
{\left(\sum\limits_{l=1}^{N}\tilde{\lambda_{i,l}}-\left(\phi_{i,j}+\delta_{i,j}\right)\tilde{\lambda_{i,j}}^2-\left[\delta_{i,j}\left(\sum\limits_{k\ne j,k\in \mathbb{N}}\tilde{\lambda_{i,k}}\right)+\nu_{i,j}+\mu_{i,j}\right]\tilde{\lambda_{i,j}}-\nu_{i,j}\left(\sum\limits_{k\ne j,k\in \mathbb{N}}\tilde{\lambda_{i,k}}\right)\right)^2}
}
\nonumber\\
&\times&\left(\left(\theta_{i,j}+\alpha_{i,j}\right)\tilde{\lambda_{i,j}}^2+\left[\alpha_{i,j}\left(\sum\limits_{k\ne j,k\in \mathbb{N}}\tilde{\lambda_{i,k}}\right)+\gamma_{i,j}+\beta_{i,j}\right]\tilde{\lambda_{i,j}}+\gamma_{i,j}\left(\sum\limits_{k\ne j,k\in \mathbb{N}}\tilde{\lambda_{i,k}}\right)\right)\nonumber\\
&+&\frac{2\left(\theta_{i,j}+\alpha_{i,j}\right)\tilde{\lambda_{i,j}}+\alpha_{i,j}\left(\sum\limits_{k\ne j,k\in \mathbb{N}}\tilde{\lambda_{i,k}}\right)+\gamma_{i,j}+\beta_{i,j}}
{\sum\limits_{l=1}^{N}\tilde{\lambda_{i,l}}-\left(\phi_{i,j}+\delta_{i,j}\right)\tilde{\lambda_{i,j}}^2-\left[\delta_{i,j}\left(\sum\limits_{k\ne j,k\in \mathbb{N}}\tilde{\lambda_{i,k}}\right)+\nu_{i,j}+\mu_{i,j}\right]\tilde{\lambda_{i,j}}-\nu_{i,j}\left(\sum\limits_{k\ne j,k\in \mathbb{N}}\tilde{\lambda_{i,k}}\right)}\label{partial H_ij}
\end{eqnarray}
\end{figure*}
The processor utilization constraint can be written as (\ref{r.s. lamda_i,j equivalent Processor utilization constraint}). Thus
\begin{figure*}
\begin{eqnarray}\label{r.s. lamda_i,j equivalent Processor utilization constraint}
\left(\phi_{i,j}+\delta_{i,j}\right)\tilde{\lambda_{i,j}}^2+\left[\delta_{i,j}\left(\sum\limits_{k\ne j,k\in \mathbb{N}}\tilde{\lambda_{i,k}}\right)+\nu_{i,j}-1+\mu_{i,j}\right]\tilde{\lambda_{i,j}}+(\nu_{i,j}-1)\left(\sum\limits_{k\ne j,k\in \mathbb{N}}\tilde{\lambda_{i,k}}\right)<0, \forall j\in\mathbb{N}
\end{eqnarray}
\end{figure*}
\begin{eqnarray}
\frac{\partial{(\zeta_j-1)}}{\partial{\tilde{\lambda_{i,j}}}}&=&2\left(\phi_{i,j}+\delta_{i,j}\right)\tilde{\lambda_{i,j}}
\nonumber\\
&+&\delta_{i,j}\left(\sum\limits_{k\ne j,k\in \mathbb{N}}\tilde{\lambda_{i,k}}\right)+\nu_{i,j}-1+\mu_{i,j}.~~~~
\end{eqnarray}
Then we arrive at (\ref{KKT Condition}), which completes the proof.

\section{Proof of Lemma \ref{Lemma:KKT Conditions for suboptimization}}\label{Appendix:suboptimization KKT proof}
For problem $\mathrm{(P_3)}$, the KKT conditions are
\begin{subequations}
\begin{numcases}{}
\nabla T_i^{*}(t_i)-\sum\limits_{j}\epsilon_j\nabla (\tilde{\lambda_{i,j}})+\chi\nabla(\tilde{\lambda}_{i}-\lambda_i)+\nonumber\\
\sum\limits_{j}\varpi_j\nabla(\zeta_j-1|t_i)+\rho\nabla(P_{i}-\mathcal{E}_i- \mathfrak{C}_i)=0,~\label{KKT1 of suboptimization}\\
  \tilde{\lambda_{i,j}}\ge 0,\forall j\\
\epsilon_j\ge0,\forall j\\
\epsilon_j\tilde{\lambda_{i,j}}=0,\forall j\\
  \tilde{\lambda}_{i}-t_i=0,\label{}\\
  \zeta_j-1\le0,\label{} \forall j~~~~~~~~\\
  \varpi_j\ge0,\forall j\\
  \varpi_j\Big(\zeta_j-1\Big)=0,\forall j~~~~~~\\
  P_{i}-\mathcal{E}_i- \mathfrak{C}_i\le 0,\\
  \rho\ge 0, \\
  \rho(P_{i}-\mathcal{E}_i- \mathfrak{C}_i)=0. \label{}
\end{numcases}
\end{subequations}
(\ref{KKT1 of suboptimization}) can be derived as
\begin{eqnarray}\label{kkt1 of P3}
\lefteqn{
\frac{\partial{T_i^{*}(t_i)}}{\partial{\tilde{\lambda_{i,j}}}}-\epsilon_j+\chi+\varpi_j \Big(2\frac{\phi_{i,j}}{t_i}\tilde{\lambda_{i,j}}+\delta_{i,j}+\frac{\mu_{i,j}}{t_i}\Big)
}\nonumber\\
&+&\rho\left(P_{i,j}\frac{\bar{A_{i,j}}}{R_{i,j}}+\frac{\bar{r_i}}{f_i}\eta_if_{i}^3\right)=0, \forall j.
\end{eqnarray}
The
partial derivative is
\begin{eqnarray}\label{grad of T_i^*}
\lefteqn{
\frac{\partial{T_i^{*}(t_i)}}{\partial{\tilde{\lambda_{i,j}}}}=
\phi_{i,j}
+\frac{1}{2}\Bigg[\bigg(3\theta_{i,j}\tilde{\lambda_{i,j}}^2+2(\beta_{i,j}+\alpha_{i,j}t_i)\tilde{\lambda_{i,j}}
}\nonumber\\
&+&
\gamma_{i,j}t_i\bigg)\left(-\phi_{i,j}\tilde{\lambda_{i,j}}^2-(\mu_{i,j}+\delta_{i,j}t_i)\tilde{\lambda_{i,j}}+t_i-\nu_{i,j}t_i\right)\nonumber\\
&+&\bigg(\theta_{i,j}\tilde{\lambda_{i,j}}^3+(\beta_{i,j}+\alpha_{i,j}t_i)\tilde{\lambda_{i,j}}^2+\gamma_{i,j}t_i\tilde{\lambda_{i,j}}\bigg)\nonumber\\
&\times&\bigg(-2\phi_{i,j}\tilde{\lambda_{i,j}}-(\mu_{i,j}+\delta_{i,j}t_i)\bigg)\Bigg]\nonumber\\
&\times&\left(-\phi_{i,j}\tilde{\lambda_{i,j}}^2-(\mu_{i,j}+\delta_{i,j}t_i)\tilde{\lambda_{i,j}}+t_i-\nu_{i,j}t_i\right)^{-2}.
 \end{eqnarray}
After substituting (\ref{grad of T_i^*}) in to (\ref{kkt1 of P3}) and some manipulations, we have (\ref{Partial for Suboptimization}).
Along with (\ref{Conzeta}) and (\ref{ConP}), we reach (\ref{KKT Conditions for suboptimization}).

\ifCLASSOPTIONcaptionsoff
  \newpage
\fi




\begin{thebibliography}{1}



\bibitem{N. Abbas Y. Zhang A. Taherkordi and T. Skeie201802:IEEETTJ}
N. Abbas, Y. Zhang, A. Taherkordi, and T. Skeie,~\lq\lq Mobile edge computing: A survey, \rq\rq~ \emph{IEEE Internet Things J.}, vol. 5, no. 1, pp. 450-465, Feb. 2018
\bibitem{L. LIN X. LIAO H. JIN and P. LI201908:Proc. IEEE}
L. Lin, X. Liao, H. Jin and P. Li, ~\lq\lq Computation offloading toward edge computing,\rq\rq~ \emph{Proc. IEEE}, vol. 107, no. 8, pp. 1584-1607, Aug. 2019.
\bibitem{T Zhang 201802 IEEEAccess}
T. Zhang,~\lq\lq Data offloading in mobile edge computing: A coalition and pricing based approach,\rq\rq~ \emph{ IEEE Access}, vol. 6, no. 1, pp. 2760- 2767, Feb. 2018

\bibitem{Keqin Li:IEEETSUSC}
K. Li, ~\lq\lq Computation offloading strategy optimization
with multiple heterogeneous servers in mobile
edge computing,\rq\rq~ \emph{IEEE Trans. Sustain. Comput.}, doi:10.1109/TSUSC.2019.2904680, to appear
\bibitem{L. Dong M. N. Satpute J. Shan B. Liu Y. Yu T. Yan201907:IEEEICDCS}
L. Dong, M. N. Satpute, J. Shan, B. Liu, Y. Yu, and T. Yan,~\lq\lq Computation offloading for mobile-edge computing with multi-user, \rq\rq~ \emph{Proc. IEEE ICDCS}, Dallas, TX, USA, 7-10 July 2019


\bibitem{J. L. D. Neto201811:IEEETMC}
J. L. D. Neto, S.-Y. Yu, D. F. Macedo, J. M. S. Nogueira, R. Langar, and S. Secci,~\lq\lq ULOOF: A user level online offloading framework for mobile edge computing,\rq\rq~ \emph{IEEE Trans. Mobile Comput.}, vol. 17, no. 11, pp.2660 - 2674, Nov. 2018

\bibitem{X. Chen L. Jiao W. Li X. Fu201610:IEEEToN}
X. Chen, L. Jiao, W. Li, and X. Fu,~\lq\lq Efficient multi-user computation offloading for mobile-edge cloud computing,\rq\rq~ \emph{IEEE/ACM Trans. Netw}, vol. 24, no. 5, pp. 2795 - 2808, Oct. 2016

\bibitem{S. Ulukus A. Yener E. Erkip O. Simeone M. Zorzi P. Grover and
K. Huang201501:IEEE JSAC}
S. Ulukus, A. Yener, E. Erkip, O. Simeone, M. Zorzi, P. Grover, and K. Huang, ~\lq\lq Energy harvesting wireless communications: A review of recent advances,\rq\rq~ \emph{IEEE J. Sel. Areas Commun.},
vol. 33, no. 3, pp. 360-381, Jan.2015.
\bibitem{T. Zhang201504:IEEE Trans. Veh. Technol.}
T. Zhang, W. Chen, Z. Han, and Z. Cao, ~\lq\lq A cross-layer perspective on energy harvesting aided green communications over fading channels,\rq\rq~ \emph{IEEE Trans. Veh. Technol.}, vol. 64, no. 4, pp. 1519-1534, Apr. 2015.
\bibitem{H. Azarhava J. M. Niya:IEEE WCL}
H. Azarhava, and J. M. Niya,~\lq\lq Energy efficient resource allocation in wireless energy harvesting sensor networks,\rq\rq~ \emph{IEEE Wireless Commun. Lett.}, doi: 10.1109/LWC.2020.2978049, to appear.


\bibitem{D. Altinel and G. K. Kurt201906:IEEE Transactions on Green Communications and Networking}
D. Altinel, and G. K. Kurt,~\lq\lq Modeling of hybrid energy harvesting communication systems,\rq\rq~ \emph{IEEE Trans. Green Commun. Netw.}, vol. 3, no. 2, pp. 523 - 534, Jun. 2019

\bibitem{Y. Alsaba2018:IEEE Communications Surveys Tutorials}
Y. Alsaba, S. K. A. Rahim, and C. Y. Leow,~\lq\lq Beamforming in wireless energy harvesting communications systems: A survey,\rq\rq~ \emph{IEEE Commun. Surv. Tuts.}, vol. 20 , no. 2, pp. 1329 - 1360, Secondquarter 2018



\bibitem{Y. Mao J. Zhang and K. B. Letaief201612:IEEE JSAC}
Y. Mao, J. Zhang, and K. B. Letaief, ~\lq\lq Dynamic computation offloading for mobile-edge computing with energy harvesting devices,\rq\rq~ \emph{IEEE J. Sel. Areas Commun.}, vol. 34, no. 12, pp. 3590-3605, Dec. 2016.

\bibitem{J. Xu L. Chen and S. Ren201707:IEEE TCCN}
J. Xu, L. Chen, and S. Ren, ~\lq\lq Online learning for offloading and autoscaling in energy harvesting mobile edge computing,\rq\rq~ \emph{IEEE Tran. Cogn. Commun. and Netw.}, vol. 3, no. 3, pp. 361 - 373, Jul. 2017
\bibitem{Z. Wei201906:IEEE itj}
Z. Wei, B. Zhao, J. Su, and X. Lu,~\lq\lq Dynamic edge computation offloading for internet of things with energy harvesting: A learning method,\rq\rq~ \emph{IEEE Internet Things J.}, vol. 6, no. 3, pp.4436 - 4447, June 2019

\bibitem{W. Chen D. Wang and K. Li201909:IEEE TSC}
W. Chen, D. Wang, and K. Li,~\lq\lq Multi-user multi-task computation offloading in green mobile edge cloud computing,\rq\rq~ \emph{IEEE Trans. Serv. Comput.}, vol. 12, no. 5, pp. 726 - 738, Sept.-Oct. 2019

\bibitem{D. Zeng201911:IEEE netw}
D. Zeng, S. Pan, Z. Chen, and L. Gu, ~\lq\lq An MDP-based wireless energy harvesting decision strategy for mobile device in edge computing,\rq\rq~ \emph{IEEE Network}, vol. 33, no. 6, pp. 109-115, Nov.-Dec. 2019

\bibitem{D. Zhang202004:IEEE TMC}
D. Zhang, L. Tan, J. Ren, M. K. Awad, S. Zhang, and Y. Zhang,~\lq\lq Near-optimal and truthful online auction for computation offloading in green edge-computing systems,\rq\rq~ \emph{IEEE Trans. Mobile Comput.}, vol. 19, no. 4, pp. 880 - 893, Apr. 2020.

\bibitem{F. Zhou R. Q.Hu:IEEE TWC}
F. Zhou, and R. Q. Hu,~\lq\lq Computation efficiency maximization in wireless-powered mobile edge computing networks,\rq\rq~ \emph{IEEE Trans. Wireless Commun.}, doi:10.1109/TWC.2020.2970920, to appear

\bibitem{L. Huang S. Bi and Y.-J. A. Zhang:IEEE TMC}
L. Huang, S. Bi, and Y.-J. A. Zhang, ~\lq\lq Deep reinforcement learning for online computation offloading in wireless powered mobile-edge computing networks,\rq\rq~ \emph{IEEE Trans. Mobile Comput.}, doi:10.1109/TMC.2019.2928811, to appear.





\bibitem{Kleinrock1975:Queueing Systems Volume 1}
L.~Kleinrock, \emph{Queueing Systems, Volume 1: Theory}, John Wiley and
Sons, New York, 1975.

\bibitem{S. Guo201604:Proc. IEEE INFOCOM}
S. Guo, B. Xiao, Y. Yang, and Y. Yang,~\lq\lq Energy-efficient dynamic offloading and resource scheduling in mobile cloud computing,\rq\rq~ \emph{Proc. IEEE INFOCOM}, San Francisco, CA, USA, Apr.10-15, 2016, pp. 1-9.

\bibitem{I. L. Glicksberg1952:Proceedings of the National Academy of Sciences}
I. L. Glicksberg,~\lq\lq A further generalization of the Kakutani fixed point theorem, with application to Nash equilibrium points,\rq\rq~ \emph{Proc. Amer. Math. Soc.}, vol. 3, pp. 170-174, 1952.



\end{thebibliography}
%

%
\newpage

\begin{IEEEbiographynophoto}{Tian Zhang}
received the B.S. and M.S. degrees from Shandong Normal University, Jinan, China, in 2006, and 2009, respectively, and the Ph.D. degree from Shandong University, Jinan, China, in 2014. From 2010 to 2013, he was a visiting Ph.D. student at Tsinghua University. From Sept. 2014 to Sept. 2019, he was with Shandong Normal University. Since Sept. 2019, he has been a Full Professor at Shandong Management University, also the Director of Intelligent Optimization \& Communications Research Center. His research interests include wireless communications and intelligent optimization. Dr. Tian Zhang was the recipient of the 2018 Shandong Provincial-level Teaching Achievement Award (1st class), the 2016 Shandong Province Higher Educational Science and Technology Award (3rd class), the 2015 Excellent Doctoral Dissertation Award of Shandong University, and the 2010 Science and Technology Progress Award of Shandong Province (2nd class). He served/serves as TPC members for IEEE GLOBECOM 2017, 2018, IEEE VTC-Spring 2018, IEEE 5G WORLD FORUM 2018, 2019, 2020.
\end{IEEEbiographynophoto}

\begin{IEEEbiographynophoto}{Wei Chen}
(S'05-M'07-SM'13) received the B.S. and Ph.D. degrees (Hons.) from Tsinghua University in 2002 and 2007, respectively. From 2005 to 2007, he was a Visiting Ph.D. Student with The Hong Kong University of Science and Technology. Since 2007, he has been on the Faculty at Tsinghua University, where he is currently a tenured Full Professor, also the Director of degree office of Tsinghua University, and also a university council member. From 2014 to 2016, he served as the Deputy Head of the Department of Electronic Engineering. He visited Princeton University, Telecom ParisTech, and the University of Southampton in 2016, 2014, and 2010, respectively. His research interests are in the areas of communication theory, stochastic optimization, and statistical learning.

He is a Cheung Kong Young Scholar and a member of national program for special support for eminent professionals, also known as 10,000-talent program. He has also been supported by the National 973 Youth Project, the NSFC Excellent Young Investigator Project, the New Century Talent Program of the Ministry of Education, and the Beijing Nova Program. He received the IEEE Marconi Prize Paper Award and the IEEE Comsoc Asia Pacific Board Best Young Researcher Award in 2009 and 2011, respectively. He was a recipient of the National May 1st Labor Medal and the China Youth May 4th Medal. He has served as a TPC Co-Chair for IEEE VTC-Spring, in 2011, and a Symposium Co-Chairs for IEEE ICC and GLOBECOM. He serves as an Editor for the IEEE TRANSACTIONS ON COMMUNICATIONS.
\end{IEEEbiographynophoto}

%
%

%




\end{document}